\documentclass{article}
\usepackage{amscd,verbatim}
\usepackage[all]{xy}
\usepackage{graphicx}

\usepackage{amsmath}
\usepackage[psamsfonts]{amssymb}
\usepackage{amsthm}
\usepackage{euscript}

\usepackage{xypic}
\input{xypic}

\usepackage{epsfig}
\usepackage{amssymb,amsmath,amsthm,amscd}

\usepackage{latexsym}

\renewcommand{\(}{\begin{equation}}
\renewcommand{\)}{end{equation} \vspace{-.05in}\linebreak}

\newcounter{saveeqn}
\newcounter{savealpheqn}

\newcommand{\alpheqn}{\setcounter{saveeqn}{\value{equation}}%
  \stepcounter{saveeqn}\setcounter{equation}{0}%
  \renewcommand{\theequation}{\mbox{\arabic{section}.\arabic{saveeqn}
\alph{equation}}}
  \renewcommand{\)}{\end{equation}}}
\def\part#1{\frac{\partial}{\partial{#1}}}%
\def\group#1{\refstepcounter{equation}\setcounter{saveeqn}{\value{equati
on}}%
  \label{#1}\setcounter{equation}{0}%
\renewcommand{\theequation}{\mbox{\arabic{section}.\arabic{saveeqn}
\alph{equation}}}
  \renewcommand{\)}{\end{equation}}}
\newcommand{\reseteqn}{\setcounter{equation}{\value{saveeqn}}%
  \renewcommand{\theequation}{\arabic{section}.\arabic{equation}}%
  \renewcommand{\)}{\end{equation}}}

\newcommand{\aalpheqn}{\setcounter{saveeqn}{\value{equation}}%
  \stepcounter{saveeqn}\setcounter{equation}{0}%
  \renewcommand{\theequation}{\mbox{
        \Alph{subsection}.\arabic{saveeqn}\alph{equation}}}
   \renewcommand{\)}{\end{equation}}}
\newcommand{\areseteqn}{\setcounter{equation}{\value{saveeqn}}%
  \renewcommand{\theequation}{\Alph{subsection}.\arabic{equation}}%
  \renewcommand{\)}{\end{equation}}}

\renewcommand{\thefootnote}{\alph{footnote}}
\renewcommand{\(}{\begin{equation}}
\renewcommand{\)}{\end{equation}}
\newcommand{\ba}{\begin{eqnarray}}
\newcommand{\ea}{\end{eqnarray}}

\newcommand{\bp}{\mathop{\vtop{\ialign{##\crcr
   $\hfil\displaystyle{}\hfil$\crcr\noalign{\kern-13pt\nointerlineskip}
   \BIG{(}\hskip0pt\crcr\noalign{\kern3pt}}}}}
\newcommand{\cbp}{\mathop{\vtop{\ialign{##\crcr
   $\hfil\displaystyle{}\hfil$\crcr\noalign{\kern-13pt\nointerlineskip}
   \BIG{)}\hskip0pt\crcr\noalign{\kern3pt}}}}}
\newcommand{\pa}{\mathop{\vtop{\ialign{##\crcr

$\hfil\displaystyle{\oplus}\hfil$\crcr\noalign{\kern+1pt\nointerlineskip
}
   \hspace{.08in}$^{\alpha=0}$\hskip6pt\crcr\noalign{\kern3pt}}}}}

\newcommand{\R}{\ensuremath{\mathbb R}}

\newcommand{\C}{\ensuremath{\mathbb C}}

\newcommand{\Q}{\ensuremath{\mathbb Q}}

\newcommand{\Z}{\ensuremath{\mathbb Z}}

\def\dwn{\downarrow}

\def\H{\ensuremath{{\mathbb H}}}

\newcommand{\beq}{\begin{equation}}
\newcommand{\eeq}{\end{equation}}

\newcommand{\bb}[1]{\ensuremath{\mathbb{#1}}}

\numberwithin{equation}{section}

\def\thru#1{\mathrel{\mathop{#1\!\!\!\!/}}}

\def\hsp#1{\hspace{#1in}}

\catcode`\@=11
\def\vereq#1#2{\lower3pt\vbox{\baselineskip1.5pt \lineskip1.5pt
\ialign{$\m@th#1\hfill##\hfil$\crcr#2\crcr\sim\crcr}}}
\catcode`\@=12

\makeatletter
\newcommand\figcaption{\def\@captype{figure}\caption}
\newcommand\tabcaption{\def\@captype{table}\caption}
\makeatother
\renewcommand{\(}{\begin{equation}}
\renewcommand{\)}{\end{equation}}

\newcommand{\bea}{\begin{eqnarray}}
\newcommand{\eea}{\end{eqnarray}}

\newcommand{\CC}{{\mathbb C}}

\theoremstyle{plain}
\newtheorem{theorem}{Theorem}[section]
\newtheorem{lemma}[theorem]{Lemma}
\newtheorem{prop}[theorem]{Proposition}

\theoremstyle{definition}

\newcommand{\CP}{\CC \text{P}}

\begin{document}

\begin{titlepage}


\vspace{2em}
\def\thefootnote{\fnsymbol{footnote}}

\begin{center}
{\Large\bf On the geometry of the supermultiplet in M-theory} 
\end{center}
\vspace{1em}

\begin{center}
\Large  Hisham Sati
\footnote{E-mail:
{\tt hisham.sati@yale.edu}}
\end{center}

\begin{center}
\vspace{1em}
{\em { Department of Mathematics\\
Yale University\\
New Haven, CT 06511}}\\
\vspace{3mm}
{\em {
Department of Mathematics \\
University of Maryland\\
College Park, MD 20742
}}\\

\end{center}

\vspace{0.5cm}
\begin{abstract}
The massless supermultiplet of eleven-dimensional 
supergravity can be generated from the
decomposition of certain representation of the exceptional Lie group $F_4$ into those 
of its 
maximal compact subgroup 
${\rm Spin}(9)$. In an earlier paper, a dynamical Kaluza-Klein 
origin of this observation is proposed with internal space the
 Cayley plane, ${\bb O} P^2$, and topological aspects are
 explored. In this paper we consider the
 geometric aspects and characterize the corresponding 
 forms which contribute to the action as well as cohomology 
 classes, including torsion, which contribute to the partition function.
 This involves constructions with bilinear forms.
The compatibility with various string theories are discussed,
including reduction to loop bundles in ten dimensions.
 
 \end{abstract}
 
 \end{titlepage}
 
 
 \tableofcontents
 
\section{Introduction}

We propose an origin of the massless multiplet in M-theory 
as Cayley plane bundles $\mathbb{O}P^2$ over eleven-dimensional
spacetime.  This is a continuation of the paper \cite{OP2}, where topological 
and number-theoretic aspects were explored. In this paper we focus on 
the geometric aspects and discuss some implications on physical
constructs, such as the partition function and supersymmetry.  

\vspace{3mm}
The eleven-dimensional massless supermultiplet $(g, C_3, \Psi)$, composed of the 
metric $g$, the $C$-field $C_3$, and the Rarita-Schwinger field
$\Psi$, is related to $F_4$, the exceptional Lie group of rank 4. 
Ramond \cite{Ram1} \cite{Ram2} \cite{Ram3} 
gave evidence for $F_4$ coming from the following two related
observations:

{\bf 1.}  $F_4$ appears explicitly  \cite{Ram3} in the light-cone formulation of 
supergravity in eleven dimensions \cite{CJS}. 
The generators $T^{\mu \nu}$ of the little group $SO(9)$ of the Poincar\'e
group $ISO(1,10)$ in eleven dimensions and the spinor 
generators $T^a$ combine to form the 52 operators that 
generate the exceptional Lie algebra $\frak{f}_4$ such that 
the constants $f^{\mu \nu ab}$ in the commutation relation
\(
[T^{\mu \nu}, T^a]= if^{\mu \nu ab} T^b
\label{tij}
\)
are the structure constants of $\frak{f}_4$.
The 36 generators $T^{\mu \nu}$ are in the adjoint of $SO(9)$ and
the 16 $T^a$ generate its spinor representation. This can be viewed as
the analog of the construction of $E_8$ out of the 
generators of $SO(16)$ and of $E_8/SO(16)$ in \cite{GSW}.

{\bf 2.} The identity representation of $F_4$, i.e. the one corresponding
 to Dynkin index $[0,0,0,0]$, generates
the three representations of ${\rm Spin}(9)$ \cite{Ram1} 
$
{\rm Id}(F_4) \longrightarrow (44, 128, 84)
$, 
the numbers on the right hand side correctly matching the 
number of degrees of freedoms of 
the massless bosonic content of eleven-dimensional supergravity with the individual
summands corresponding, respectively, to the graviton, the
gravitino, and the $C$-field (see the beginning of section \ref{count}).

\vspace{3mm}
The main Idea of this paper, presented in section \ref{sec;euler}
is interpreting Ramond's triplets as arising from 
${\bb O} P^2$ bundles with structure group $F_4$ over our 
eleven-dimensional manifold $Y^{11}$, on which M-theory is 
`defined'. We first discuss in section \ref{sec:spin9}
the geometric properties of $\mathbb{O}P^2$, including 
${\rm Spin}(9)$-structures and characteristic classes. This 
leads to one of the main results, theorem \ref{thm:fields}, 
that the massless fields of M-theory are encoded in the spinor bundle 
of ${\bb O} P^2$. We then relate ${\rm Spin}(9)$-structures on 
the 9-dimensional vector space $V^9$ to the geometry of
the eight-sphere $S^8$ which in turn, by \cite{JFO}, is related
to Killing spinors on the cone over $S^8$.  This shows that the
unification of the fields, as well as their supersymmetry, in M-theory 
can be seen from the eight-sphere over the Cayley plane 
(cf. proposition \ref{prop:f4}). We then show that 
fields can be given yet another interpretation via the index 
of the twisted (Kostant) Dirac operator of
 \cite{Land1}. 
 The identity representation of $F_4$ encoding the supergravity multiplet is 
 the space of twisted harmonic spinors on $\mathbb{O} P^2$, which is
 propositon 
\ref{thm:harmonic}.
 
 \vspace{3mm}
 After studying structures on $\mathbb{O}P^2$, we use that space itself 
 as the fiber over eleven-dimensional spacetime. In section 
\ref{geom conseq} we explore the consequences of this idea by 
relating the geometry and the characteristic classes of the 
base to that of the total space, using the knowledge of that 
of the fiber studied in section \ref{count}.  If the base 
$Y^{11}$ has positive Ricci curvature then so does the total
space $M^{27}$. This and related matters are discussed in
section \ref{geom conseq}.
In section \ref{strs} we relate structures, such as Fivebrane structures
\cite{SSS1} \cite{SSS2}, as well as 
genera on the base space to genera on the total 
space. This includes elliptic genera, Witten genera, Ochanine genera 
and is the content of proposition \ref{ellm27}. We use this to 
relate the genera to an elliptic refinement of the mod index of the 
Dirac operator appearing in the study of the M-theory partition 
function \cite{DMW} \cite{KS1}.

%





%


\vspace{3mm}
In section \ref{sec:terms} we consider possible terms in the lifted
action up in twenty-seven dimensions. In particular, in section
\ref{sec:cayley} we consider 
the Cayley 8-form, which is a generalization to manifolds of 
${\rm Spin}(9)$ holonomy of the Cayley 4-form or the K\"ahler 
2-form on manifolds with quaternionic and complex structures,
and identify that 8-form as a representative in the cohomology 
of the Cayley plane and as a possible term in the lifted action.
Then in section \ref{effect on PF}
we consider torsion classes and their effect on the M-theory
partition function.
We show in propositions \ref{thm:comp} and \ref{thm:cancel}
that $\Z_2$ and $\Z_3$ classes from the classifying space $BF_4$ are 
compatible with the M-theory 
partition function.

\vspace{3mm} 
In  section \ref{sec:further} we consider further possible terms
in twenty-seven dimensions. In particular, in section 
\ref{sec:kin}
we consider possible terms, lifting 
the degree eight class introduced in \cite{DFM},
and which generalize the $G_4\wedge*G_4$ term 
in the eleven-dimensional action. A natural question then
arises whether the construction in this paper is compatible
with type II superstring theory in ten dimensions and 
bosonic string theory in twenty-six dimensions.
We study the former in section \ref{sec:10}, where we show
that the dimensional reduction of the $F_4$ bundle on the circle in $Y^{11}$ leads to 
an $LF_4$ bundle over $X^{10}$ and, under some natural assumptions, 
compatibility 
with type II string theory. We discuss the latter, i.e. the compatibility with 
bosonic string theory, in section \ref{compat}. 
Finally we collect in the appendix some basic useful properties of the 
Cayley plane.

\vspace{3mm}
We use the Lorentz signature in studying the spectrum in section \ref{count},
and then resort to the Euclidean signature when discussing the geometric 
aspects in the rest of the paper. 

\section{The Fields in M-theory}
\label{count}
The low energy limit of M-theory (cf. \cite{Dynamics} \cite{Town}
\cite{Duff2}) is eleven-dimensional supergravity \cite{CJS}, whose 
field content on an eleven-dimensional spin manifold $Y^{11}$ with Spin 
bundle $SY^{11}$ is 
\begin{itemize}
\item {\underline{Two bosonic fields}}: The metric $g$ and the three-form $C_3$. It is 
often convenient to work with Cartan's moving frame formalism so that the metric is 
replaced by the 11-bein $e_M^A$ such that $e_M^A e_N^B= g_{MN} \eta^{AB}$,
where $\eta$ is the flat metric on the tangent space.
\item {\underline{One fermionic field}}: The Rarita-Schwinger vector-spinor $\Psi_{1}$,
which is classically a section of $SY^{11} \otimes TY^{11}$, i.e. a spinor 
coupled to the tangent bundle. 
\end{itemize}

The count of the on-shell degrees of freedom, i.e. components, of the fields
is done by eliminating the redundant gauge degrees of freedom. This could be
done for example by choosing the light cone gauge: decompose Minkowski space
$\R^{1,10}$ into $\R^{1,1} \oplus \R^9$, with 
$\R^{1,1}={\rm Span}({\bf v}_1, {\bf v}_2)$
where the vectors ${\bf v}_i$ satisfy 
$|{\bf v}_1|^2=|{\bf v}_2|^2=0$ and ${\bf v}_1 \cdot {\bf v}_2\neq 0$.

\vspace{3mm}
The Poincar\'e group $\R^{1, 10} \ltimes SO(1, 10)$ corresponds to the 
algebra $\R^{1, 10} ~\widetilde{\oplus} ~\frak{so}(1,10)$ where the brackets
$[\R^{1, 10}, \frak{so} (1.10)]$ are given by the vector representation of 
$\frak{so} (1,10)$ on $\R^{1, 10}$.  Since the latter is abelian then the 
irreducible representations are one-dimensional, and hence given by
the characters $(\R^{1, 10})^*$. This is acted upon by $\frak{so} (1,10)$,
which decomposes the space of characters into orbits characterized by
the mass $m^2=|{\bf v}|^2$ for ${\bf v} \in (\R^{1, 10})^*$. Let $H$ be the stabilizer 
of a point. $H$ is called the little group. An irreducible representation 
of the Poincar\'e algebra is the space of sections of a homogeneous vector 
bundle $E= SO(1,10) \times_H K$ over the orbit $SO(1, 10)/H$, where
$K$ is a representation of $H$. The representations, by the Wigner classification,  
are as follows:
\begin{itemize}
\item {\it Massive fields}: For $|{\bf v}|^2 \neq 0$ the little group is $H=SO(10)$.
\item {\it Massless fields}: For $|{\bf v}|^2=0$ the little group is $H=SO(9)$.
\end{itemize}

The states for eleven-dimensional supergravity are massless and hence
 form irreducible representations of the little group  $SO(9)$. 
The count is is as follows (with $D=11$ ):
\begin{enumerate}
\item {\it The 11-bein $e_M^A$}: Traceless symmetric $(D-2) \times (D-2)$ matrix gives
$\frac{1}{2}D(D-3)=44$ \cite{Julia1}.
\item {\it The $C$-field $C_3$}: A 3-form in $\R^9$ gives 
$\binom{D-2}{3}=\frac{(D-2)!}{3!(D-2-3)!}=84$.
\item {\it The Rarita-Schwinger field $\Psi_1$}: $2^{\frac{1}{2}(D-1)-1}(D-3)=128$,
where the factor of $-1$ in the exponent comes from the fact that $\Psi_1$ is 
a Majorana, i.e. real, fermion. 
\end{enumerate}

\subsection{The Euler Triplet}
\label{sec;euler}

In this section we review Ramond's observation we mentioned in the introduction 
and state the main theme of this paper. We will basically `geometrize' and 
`topologize' the representation-theoretic observation, hence making room 
for dynamics from kinematics. Therefore, the appearance of the $F_4$ 
representation and the decomposition under the maximal compact subgroup
${\rm Spin}(9)$ to give the degrees of freedom of the fields will be taken to 
originate from an ${\bb O} P^2$ bundle over $Y^{11}$. 

\vspace{3mm}
There are anomalous embeddings of certain groups into 
an orthogonal group in which the
vector representation of the bigger group is identified with the spinor
of the smaller group. For example, for $SO(9)$ we have \cite{Julia1}
\bea
SO(16) &\supset& SO(9)
\nonumber\\
{\rm vector} &=& {\rm spinor},
\label{vs}
\eea
both of dimension 16. In fact this explains the emergence of supersymmetry
for the supermultiplet of eleven-dimensional supergravity \cite{Julia1}
\cite{Duff1} \cite{Ram1}. Furthermore, in \cite{Duff1} it was conjectured that
$SO(16)$ is a local symmetry of 11-dimensional supergravity. This was proved
in \cite{Nic1}. 
One of the goals in this paper will be to  
seek a geometric origin for the above observation (eqn. (\ref{vs}) )
via ${\bb O} P^2$ bundles,
as ${\rm Spin}(16)$ will be the Spin group of the projective plane fiber.
We hope this would also shed some light on the enlarged local symmetry 
in the theory since the symmetry groups coming from bundles on 
${\bb O} P^2$ will act locally (at least on the space itself).

\vspace{3mm}
Since ${\rm rank}(F_4)={\rm rank}({\rm Spin}(9))$ then ${\bb O} P^2$ is an equal rank
symmetric space. A generalization to homogeneous spaces of the Weyl character
formula, with maximal torus replaced by the equal rank maximal compact subgroup,
is the Gross-Kostant-Ramond-Sternberg character
formula \cite{GKRS}
\(
V_{\lambda} \otimes S^+ - V_{\lambda} \otimes S^-= \sum_c {\rm sgn} (c)
U_{c \bullet \lambda},
\label{char}
\)
which can be applied as follows \cite{Ram1} to the pair $(F_4, {\rm Spin}(9))$. 
The left hand side involves the differences of tensor products of 
representations $V_{\lambda}$ of $F_4$ with highest weight $\lambda$
written in terms of its
${\rm Spin}(9)$ subgroup, and $S^{\pm}$, the two semi-spinor 
representations of ${\rm Spin}(16)$ written in terms of its embedded
subgroup ${\rm Spin}(9)$, i.e. the spin representation associated to the 
complement of $\frak{spin}(9)={\rm Lie}({\rm Spin}(9))$ in
$\frak{f}_4={\rm Lie} (F_4)$. The right hand side involves the sum over $c$, the 
elements of the Weyl group which map the Weyl chamber of $F_4$ into that
of ${\rm Spin}(9)$. The number of such elements is three, given by the 
ratio of the orders of the Weyl groups (\ref{ratio}), i.e. the subset $C \in W_{F_4}$ 
has one representative from each coset of $W_{{\rm Spin}(9)}$.
 $U_{c \bullet \lambda}$
denotes the ${\rm Spin}(9)$ representation with highest weight 
$c \bullet \lambda=c(\lambda + \rho_{F_4}) - \rho_{{\rm Spin}(9)}$, with 
$\rho$ the sum of fundamental weights. For $F_4$, as mentioned above,
there corresponds three  equivalent ways of embedding ${\rm Spin}(9)$ into
$F_4$. This implies that for each representation of $F_4$, there are
$\chi(F_4/{\rm Spin}(9))=3$ irreducible representations of ${\rm
Spin}(9)$ generated, called the {\it Euler triplet}.

\vspace{3mm}
The consequence for eleven-dimensional supergravity is that the fields
satisfy the character formula exactly for the pair $(F_4, {\rm Spin}(9))$
\cite{Ram1}.
Under the decomposition ${\rm Spin}(16) \supset {\rm Spin}(9)$, one of the 
 semi-spinor representations, $S^+$, stays the same, $128 = 128$,
  while the other, $S^-$, decomposes as $128'=44+84$. 
  For a highest weight $\lambda=0$, one gets $c(\rho_{F_4})=\rho_{SO(9)}$
  the character formula 
  is then clearly satisfied \cite{Ram1} as
  \(
  {\rm Id} \otimes S^+ - {\rm Id} \otimes S^{-} =0,
  \)
  i.e.
  \(
  128 - (44 +  84)=128 -44 -84.
  \)
The Dynkin labels of the fields in the representation of ${\rm Spin}(9)$
are $[2000]$ for the graviton as a symmetric second rank tensor,
$[0010]$ for the 3rd rank antisymmetric tensor $C_3$, and 
$[1001]$ for the Rarita-Schwinger spinor-vector.

\vspace{3mm}
\noindent {\bf Remarks}

\noindent {\bf 1.} There is a very interesting Dirac operator whose index is not zero on ${\bb O} P^2$.
This is Kostant's cubic Dirac operator \cite{Kos}
\(
\thru {\cal K} \xi := \sum_{a=1}^{16} \Gamma^a T^a \xi=  0,
\label{kos}
\)
where $\Gamma^a$, $a, b=1,2, \cdots, 16$ are $2^8 \times 2^8$
gamma matrices that generate the Clifford algebra 
$\left\{ \Gamma^a, \Gamma^b \right\}= 2\delta^{ab}$.
Solutions of the Kostant equation (\ref{kos}) 
consists of all Euler triplets, including the supergravity multiplet \cite{Ram3}.
The right hand side of (\ref{char}) is the kernel of (\ref{kos}). We will
deal with other Dirac operators in section \ref{susy}.

\noindent {\bf 2.}  The Euler characteristic of ${\bb O} P^2$ can be calculated
as the ratio of the orders of the Weyl groups
\( \chi ({\bb O} P^2)=\chi \left( F_4/ {\rm Spin}(9) \right)=
\frac{|W(F_4)|}{|W(B_4)|}= \frac{|W(F_4)|}{\Z_2^4 \odot S_4}=
\frac{2^7.3^2}{2^4.4!}=3.
\label{ratio}
 \)
 Such a formula holds for general equal rank symmetric spaces $G/H$, 
 by a classic result of Hopf and Samelson.
 
 \vspace{3mm}
 We now give the main theme around which this paper is centered.

\vspace{3mm}
\noindent {\bf Main Idea:} {\it We interpret Ramond's triplets as arising from 
${\bb O} P^2$ bundles with structure group $F_4$ over our 
eleven-dimensional manifold $Y^{11}$, on which M-theory is 
`defined'.} 

\vspace{3mm}
\noindent We have dealt with ${\bb O} P^2$ bundles systematically and in detail in  
\cite{OP2}, so now we proceed with the geometric interpretation of the main
idea, as well as propose a geometric interpretation for the observation
(\ref{vs}).

\subsection{${\rm Spin}(9)$-structures and the M-theory fields}
\label{sec:spin9}

Before putting ${\bb O} P^2$ as a fiber, we start with just the space ${\bb O} P^2$ itself.

\subsubsection{${\rm Spin}(9)$ bundles} 

We start with the Spin structure on the Cayley plane. 
 
 \begin{lemma}
 ${\bb O} P^2$ admits a unique Spin structure.
\end{lemma}

Over the homogeneous space ${\bb O} P^2=F_4/{\rm Spin}(9)$ we always 
have the canonical ${\rm Spin}(9)$ bundle, which we call ${\wp}$.
Let $\Delta : {\rm Spin}(9) \to U(16)$ be the spinor  representation. 
We can thus form associated vector bundles with structure group 
$U(16)$ over ${\bb O} P^2$. To investigate these we should look at the K-theory
of ${\bb O} P^2$. This has been done for general equal rank symmetric spaces
$G/H$ in \cite{AH}. The group $K^1(G/H)$ is zero, whereas $K^0(G/H)$ 
is a free abelian group of rank equal to the Euler number, so that 
$K^0({\bb O} P^2)=\Z \oplus \Z \oplus \Z$. Furthermore, $K^0({\bb O} P^2)$
has no torsion and the Chern character map ch : $K^0({\bb O} P^2) \to 
H^{\rm even} ({\bb O} P^2;\Q)$ is injective. Since $H^*({\bb O} P^2;\Z)$ 
has no torsion, $K^0$ is isomorphic to the cohomology of
${\bb O} P^2$. Therefore, 

\begin{prop}
A complex vector bundle over ${\bb O} P^2$ is uniquely characterized by the classes
in degrees 0, 8, and 16. 
\end{prop}
  
Let ${\Re}({\rm Spin}(9))$ be the representation ring of ${\rm Spin}(9)$ and 
let $\beta : {\Re}({\rm Spin}(9)) \to K^0({\bb O} P^2)$ be the map that
assigns vector bundles over ${\bb O} P^2$ to representations of 
${\rm Spin}(9)$, so that we have the composite map
\(
\xymatrix{
{\rm Spin}(9) 
\ar[r]^{\hspace{-3mm}\Delta}
&
{\Re}({\rm Spin}(9))
\ar[r]^{\beta}
&
K^0({\bb O} P^2)
\ar[r]^{\hspace{-4mm} \rm ch}
&
H^{\rm even}({\bb O} P^2;\Q)
}\;.
\)
In fact the map $\beta$ is surjective, which can be seen as follows \cite{AH}.
Let $s_j$ be the $j$th elementary symmetric function in the $x_i^2$,
where $x_i$, $i=1,2,3,4$, are elements of the maximal torus of 
${\rm Spin}(9)$, as in \cite{BH}. Then, using
$s_2=s_2(x_1^2, x_2^2, x_3^2, x_4^2)=\sum_{i<j} x_i x_j$ and 
$s_4=s_4(x_1^2, x_2^2, x_3^2, x_4^2)=\prod_{i=1}^4 x_i^2$, 
the Chern character 
\bea
{\rm ch}(\beta \Delta) = 2^4 \prod_{i=1}^4 {\rm cosh}\left(\frac{x_i}{2}\right)
&=& {\rm rk} + \frac{s_2}{6} + {\rm higher~terms}
\nonumber\\
&=& 16 + u + {\rm higher~terms}\;,
\eea
has $u$, the generator of $H^8({\bb O} P^2;\Z)=\Z$, as a summand.  Therefore 
we have

\begin{prop}
Every complex vector bundle over ${\bb O} P^2$ is an associated vector bundle 
for the ${\rm Spin}(9)$ principal bundle ${\wp}$.
\end{prop}

\subsubsection{${\rm Spin}(9)$-structures}

Let $\frak{f}_4$ and $\frak{spin}(9)$ be the Lie algebras of $F_4$ and 
${\rm Spin}(9)$, respectively. The adjoint action of $F_4$ is given by
\(
{\rm Ad}_{F_4}: F_4 \longrightarrow {\rm Aut}_{\rm Lie} (\frak{f}_4).
\label{ad}
\)
Consider the restriction to ${\rm Spin}(9)$
\(
{\rm Ad}_{F_4, {\rm Spin(9)}} :=
{\rm Ad}_{F_4}|_{{\rm Spin}(9)}
: {\rm Spin}(9) \longrightarrow {\rm Aut}_{\rm Lie} (\frak{f}_4),
\label{rest}
\)
which is given by
$
{\rm Ad}_{F_4}|_{{\rm Spin}(9)} (k) X= {\rm Ad}_{F_4}(k) X = {\rm Ad}_{{\rm Spin}(9)}(k)$
for $X \in \frak{spin}(9)$ and $k \in {\rm Spin}(9)$. This means that 
$\frak{spin}(9)$
is an invariant subspace for the respresentation ${\rm Ad}_{F_4}|_{{\rm Spin}(9)}$
of ${\rm Spin}(9)$ in $\frak{f}_4$, and there is the factor representation
$
{\rm Ad}^{\perp} : {{\rm Spin}(9)} \longrightarrow GL(\frak{f_4}/\frak{spin}(9))$. 
The sequence 
$
0 \longrightarrow \frak{spin}(9) \longrightarrow \frak{f}_4
\longrightarrow \frak{f}_4/ \frak{spin}(9) 
\longrightarrow 0
$
is exact and ${\rm Spin}(9)$-equivariant. 
Consider the principal fiber bundle with total space $F_4$,
$
{\rm Spin}(9)
\to
F_4 
\buildrel{p}\over{\longrightarrow}
F_4/{\rm Spin}(9)
$.
Using the representations (\ref{ad}) and (\ref{rest}) we form the associated bundles
$E_1$
\(
\xymatrix{
\frak{spin}(9)
\ar[r]
&
F_4 \times_{{\rm Spin}(9)} \frak{f_4}/\frak{spin}(9) =E_1
\ar[r]^{\hspace{1cm}\pi_1}
&
F_4/{\rm Spin}(9)
}
\label{E1}
\)
 and $E_2$
 \(
\xymatrix{
{\frak{spin}}(9)
\ar[r]
&
F_4 \times_{{\rm Spin}(9)} \frak{spin}(9) =E_2
\ar[r]^{\hspace{1cm} \pi_2}
&
F_4/{\rm Spin}(9)\; ,
}
\label{E2}
\)
 respectively. Then we have the following characterization of the tangent 
 bundle of the Cayley plane.
 
 \begin{prop}
 $T({\bb O} P^2)$ is the associated vector bundle $E_1$. Furthermore, 
 $E_1 \oplus E_2$ is a trivial vector bundle.
 \end{prop}
 Results for general $G/K$ are proved in \cite{Mich}.

\vspace{3mm}
Denote by ${\mathcal F}({\bb O} P^2)$ the frame bundle of the Cayley plane 
with structure group ${\rm S}O(16)$. A ${\rm Spin}(9)$-structure is a reduction 
${\mathcal R} \subset {\mathcal F}({\bb O} P^2)$ of the ${\rm S}O(16)$-bundle
 ${\mathcal F}({\bb O} P^2)$ via the homomorphism 
 $\kappa_9 : {\rm Spin}(9) \to {\rm S}O(16)$.
A ${\rm Spin}(9)$-structure defines certain other geometric structures \cite{Fried}. In 
particular, it induces
 a 9-dimensional real, oriented Euclidean vector bundle
$V^9$ with Spin structure given by
$
V^9 := {\mathcal R} \times_{{\rm Spin}(9)} \R^9
$.

\begin{lemma}
${\bb O} P^2$ admits a ${\rm Spin}(9)$-structure.
\end{lemma}

\proof
Due to the topology of ${\bb O} P^2$, the only nontrivial cohomology,
with any coefficients, is in the top and the middle dimension
(see Appendix). Then the 
only possible obstruction to 
reducing the structure group from ${\rm Spin}(16)$ to ${\rm Spin}(9)$ is
\(
H^8\left( 
{\bb O} P^2; \pi_{8-1}
\left( \frac{{\rm Spin}(16)}
{{\rm Spin}(9)}
\right) 
\right).
\)
From the homotopy exact sequence for the fibration 
$
{\rm Spin}(9) \longrightarrow {\rm Spin}(16)
\longrightarrow  {\rm Spin}(16)/{\rm Spin}(9)
$
and the fact that the homotopy groups of ${\rm Spin}(i)$, $i=9, 16$ are
 \bea
 \hspace{-1cm}
  \pi_{{}_{3 \leq n \leq 15}}({\rm Spin}(16)) =(\Z, 0, 0, 0, \Z, \Z_2, \Z_2, 0, \Z,
 0, 0, 0, \Z \oplus \Z) &&
 \\
  \pi_{{}_{3 \leq n \leq 15}}({\rm Spin}(9)) = \left( \Z, 0, 0, 0, \Z, \Z_2 \oplus \Z_2, \Z_2 \oplus \Z_2,
  \Z_8, \Z \oplus \Z_2, \right. &&
 \nonumber\\ 
\left.
  0, 
  \Z_2, \Z_2 \oplus \Z_8, \Z \oplus Z_2 \oplus \Z_2\oplus \Z_2 \right),&&
   \eea
  we get that $\pi_7({\rm Spin}(16)/{\rm Spin}(9))=0$. Therefore, there are no 
  obstructions to reducing the structure group from ${\rm Spin}(16)$ to 
  ${\rm Spin}(9)$.
\endproof

\begin{lemma} 
(Properties of $V^9$)
\noindent {\bf 1.} {\it Spinors:} The tangent bundle $T({\bb O} P^2)$ is isomorphic to the bundle 
${\Delta_9}(V^9)$ 
of real spinors of the
vector bundle $V^9$.

\noindent {\bf 2.} {\it Stiefel-Whitney classes:} The Stiefel-Whitney classes of ${\bb O} P^2$ are related to 
the corresponding classes of $V^9$ by the formula
$
w_8({\bb O} P^2) = w_4^2(V^ 9) + w_8(V^9)$.

\noindent {\bf 3.} {\it Pontrjagin classes:} 
$
p_1(V^9)=0 = p_3(V^9)$, 
$p_2(V^9)=-p_2({\bb O} P^2)= -6u$,
$p_4(V^9)=-\frac{1}{13}p_4({\bb O} P^2)=-3u^2$.
\end{lemma}

\proof
\noindent Part ${\bf 1}$ follows from the definition. It is known 
that $\frak{f}_4= \frak{so}(9) \oplus S^+$ \cite{Adams} \cite{Baez}.
The isotropy group ${\rm Spin}(9)$ acts on the tangent space
$T_x {\bb O} P^2=\frak{f}_4/{\frak{spin}}(9)$ as a sixteen-dimensional
representation, the spinor representation $\Delta_9$ of ${\rm
Spin}(9)$.

\noindent Part ${\bf 2}$ follows from an application of the discussion in \cite{GG} for 
a general 16-manifold with ${\rm Spin}(9)$-structure. We just show how to get the
Stiefel-Whitney classes of ${\bb O} P^2$. We use the Wu classes 
$\nu_i \in H^i( {\bb O} P^2; \Z_2)$ defined by 
\(
\langle \; \nu_i \cup u \; , \; [{\bb O} P^2] \; \rangle
= \langle \; Sq^i u \; , \; [{\bb O} P^2] \; \rangle\;,
\label{nu}
\)
where $Sq$ is the Steenrod squaring cohomology operation.
Since $Sq^8 u=u^2$ then the total Wu class of ${\bb O} P^2$
is $\nu= 1+ u + u^2$, so that, by (\ref{nu}), the total Stiefel-Whitney
class is 
\(
w ({\bb O} P^2) = Sq~ \nu = 1 + u + u^2.
\label{sw}
\)
\noindent For part ${\bf 3}$ we apply theorem 2 (or corollary 3) of \cite{Fried} to the case 
of ${\bb O} P^2$ so that we have
the following (see Appendix for the characteristic classes of $\mathbb{O}P^2$): 
First
$p_1({\bb O} P^2)= 2 p_1(V^9)=0$. 

\noindent Second, $p_2({\bb O} P^2)= \frac{7}{4}(V^9) - p_2(V^9)$
so that $p_2(V^9)=-p_2({\bb O} P^2)$ since $p_1(V^9)$ is zero. 

\noindent Third, 
$p_3({\bb O} P^2)= \frac{1}{8}\left( 7 p_1^3(V^9) - 12p_1(V^9) p_2(V^9)
+ 16 P_3(V^9) \right)$, which gives that $p_3(V^9)=0$ since $p_2(V^9)=0$
and $p_3({\bb O} P^2)=0$. 

\noindent Fourth, $p_4({\bb O} P^2)= \frac{1}{128}
\left(35p_1^4(V^9) - 120p_1^2(V^9)p_2(V^9) + 400p_1(V^9)p_3(V^9)
-1664p_3(V^9)  \right)$, which gives \newline
$p_4(V^9)= - \frac{1}{13} p_4({\bb O} P^2)$
upon using $p_1(V^9)=0$. 
\endproof

\begin{lemma}
The Euler class and the fourth $L$-polynomial of ${\bb O} P^2$ are given
in terms of the Pontrjagin classes of $V^9$ as 
\(
e({\bb O} P^2) = \frac{p_2^2(V^9) - 4p_4(V^9)}{16}\;,
\qquad \quad
L_4( {\bb O} P^2) =-\frac{1}{3^4.5^2.7}\left( 19 p_2^2(V^9) + 4953p_4(V^9) \right)
\)
\end{lemma}

\proof
The formula for the Euler class follows either 
from substitution of the Pontrjagin classes of $V^9$ in terms of the 
Pontrjagin classes of ${\bb O} P^2$ in the Euler class formula of ${\bb O} P^2$ or
directly by observing that, with $p_1(V^9)=0$,  
\(
e({\bb O} P^2)=\frac{1}{256}p_1^4(V^9) - \frac{1}{32}p_1^2(V^9)p_2(V^9) + \frac{1}{16}p_2^2(V^9) - \frac{1}{4}p_4(V^9)
\) 
gives the answer. Finally, the formula for $L_4$ follows by direct substitution into  
\(
L_4 =\frac{1}{3^4. 5^2. 7}
\left( 381 p_4 - 71 p_3 p_1 - 19 p_2^2 + 22 p_2 p_1^2 - 3 p_1^4 \right),
\)
so that 
$
L_4( {\bb O} P^2)=-\frac{1}{3^4.5^2.7}\left( 19 p_2^2(V^9) + 4953p_4(V^9) \right)
$.
\endproof

\noindent {\bf Remark.}
Using $V^9$ we can recover the signature of ${\bb O} P^2$,   
$
\sigma({\bb O} P^2)= 
-\frac{1}{3} \int_{{\bb O} P^2} p_4(V^9) = \frac{1}{39} \int_{{\bb O} P^2} p_4({\bb O} P^2)$,
which is related to the Euler class by $e({\bb O} P^2) = 3 \sigma ({\bb O} P^2)$.

\subsubsection{Consequences for the M-theory fields}

One major advantage of the introduction of an ${\bb O} P^2$ bundle is that 
in this picture the bosonic fields of M-theory, namely the metric and the $C$-field, can
be unified. 

\begin{theorem}
The metric and the $C$-fields are orthogonal components of the
positive spinor bundle of ${\bb O} P^2$. 
\end{theorem}

\proof   
The spinor bundle $S^+({\bb O} P^2)$ of the Cayley plane is isomorphic to
\(
S^{+}({\bb O} P^2) = S_0^2(V^9) \oplus \Lambda^3 (V^9),
\label{s+}
\)
where $S_0^2$ denotes the space of traceless symmetric 2-tensors. 
This follows from an application of proposition 3 in \cite{Fried} 
which requires the study the 16-dimensional spin representations 
$\Delta_{16}^{\pm}$ as ${\rm Spin}(9)$-representations. 
The element $e_1\cdots e_{16}$ belongs to the subgroup 
${\widetilde{\rm Spin}}(9) \subset {\rm Spin}(16)$
and acts on
$\Delta_{16}^{\pm}$
 by multiplication by $(\pm1)$. 
This means that  $\Delta_{16}^{+}$ is an $SO(9)$-representation, 
but $\Delta_{16}^{-}$ is a ${\rm Spin}(9)$-representation \cite{Adams}.
Both representations do not
contain non-trivial ${\rm Spin}(9)$-invariant elements. Such 
an element would define a parallel spinor on 
$F_4/{\rm Spin}(9)$ but, since the Ricci tensor of ${\bb O} P^2$ is not zero 
(see section \ref{geom conseq}), the spinor must vanish by the Lichnerowicz formula 
\cite{Li}
$D^2 = \nabla^2 + \frac{1}{4} R_{\rm scal}$.
Then $\Delta_{16}^{+}$ as a ${\rm Spin}(9)$ representation is given by 
equation (\ref{s+}), and $\Delta_{16}^{-}$
is the unique irreducible Spin(9)-representation of dimension 128.
\endproof

\vspace{3mm}
\noindent {\bf Remarks}

\noindent {\bf 1.} From the above we see that the Rarita-Schwinger field is given by the negative spinor bundle 
of ${\bb O} P^2$. 

\noindent {\bf 2.} The 11-bein can also be seen from the nine-dimensional bundle in another
way. It is an element of $SL(9)/{\rm Spin}(9)$, which indeed has dimension $44$.

\noindent {\bf 3.} In \cite{KNS} it was shown that the bosonic degrees of freedom, $g$ and $C_3$,
can be assembled into an $E_{8(+8)}$-valued vielbein in eleven dimensions. As
$E_{8(+8)}$ is the global symmetry of the two factors in the symmetry group 
$E_{8(+8)} \times {\rm S}O(16)$,
it would be interesting to see whether the discussion of the second factor here might 
be related to \cite{KNS}.

\vspace{3mm}
Thus we have

\begin{theorem}
The massless fields of M-theory are encoded in the spinor bundle of ${\bb O} P^2$. 
\label{thm:fields}
\end{theorem}

Next we have the following observation

\begin{prop}
   There is no obstruction to having sections of the ${\rm Spin}(9)$ 
bundle on a manifold of dimension greater than or equal to 9.
\end{prop}

\proof
This has been observed in \cite{GG} and \cite{IPW} in a different context.
The real dimension of the spinor representation $S$ is 
$d=2^{\frac{m}{2}} \alpha$, where $\alpha$ depends on the
dimension and consequently on the condition on the spinors
(i.e. Majorana, Weyl), so that the maximum dimension $m$ 
of the manifold $M$ for which $d=m$ is $m=8$. When $m >8$
the dimensions cease to be equal anymore, ${\rm dim}~S > {\rm dim}~M$.
The obstruction bundle is the bundle of spinors of unit norm whose fiber is
$SO(d)$. As the only nontrivial homotopy group of the sphere $S^{d-1}$
in degrees less than or equal to $d-1$ is
$\pi_{d-1}(S^{d-1})=\Z$, the primary-- and only--  obstruction 
lies in $H^d(M^m; \Z)$. For $n \geq 9$ one has $d > m$, so that
the obstruction is zero.
\endproof

\paragraph{Remark.}
We can use the twisted geometric Dirac operator introduced in \cite{Land1} 
to give another interpretation of the 
the Euler triplet in M-theory. Since $\mathbb{O} P^2$ is Spin,
 the identity representation of $F_4$ is the index of the the Dirac operator 
 on $\mathbb{O} P^2$ twisted by the homogeneous vector bundle induced by the 
 representation of ${\rm Spin}(9)$. Calling this representation $\cal V$ and
 consider the representations $S_+^*$ and $S_-^*$, dual to half-Spin 
 representations $S^+$ and $S^{-}$, respectively.
 Applying \cite{Land1}, we have the twisted Dirac operator 
 \(
 D_{S(\mathbb{O} P^2) \otimes {\cal V}} : L^2 \left(F_4 \times_{{\rm Spin}(9)}
 (S_+^* \otimes {\cal V} \right)
 \longrightarrow
  L^2 \left(F_4 \times_{{\rm Spin}(9)}
 (S_{-}^* \otimes {\cal V} \right)\;,
 \label{def twi dir}
 \)
 whose index is 
 \( 
 {\rm Index}\hspace{0.5mm} D_{S(\mathbb{O} P^2) \otimes {\cal V}}
  ={\rm Id} \hspace{0.5mm} (F_4)
 \;.
\label{twisted index}
 \)

\subsection{Supersymmetry} 
\label{susy}

 We have seen that supersymmetry is created from bundles on ${\bb O} P^2$.
 More precisely, this is really due to parallel spinors on $\R^9$.  
In fact, this can be seen from another angle. There is a 
supersymmetric structure inside of $V^9$, which makes $\frak{f}_4$ into
a Lie superalgebra. The connection comes from the relation between 
real Killing spinors on a space and the parallel spinors on the cone
over that space \cite{Bar}. Let us see how this works, following \cite{JFO}.
The eight-sphere $S^8$ with the standard round metric $g$ has a Spin
bundle $S(S^8)$ on which there is an action of the Clifford bundle 
$C\ell (TS^8)$ and a Spin$(8)$ invariant inner product. A Killing spinor
over $S^8$ is a nonzero section $\epsilon$ of $S(S^8)$ which satisfies,
for all vector fields $X$,
$
\nabla_X \epsilon = \lambda X \cdot \epsilon
$,
with Killing constant $\lambda \in \R$. In local coordinates, using
$\lambda =\frac{1}{2}$, this is
\(
(\nabla_{\mu} - \frac{1}{2} \gamma_{\mu}) \epsilon=0.
\)
The cone on $S^8$ is $\mathcal{C}S^8= \R^9 \setminus \{ 0 \}$. The metric
$dr^2 + r^2 g$, however, can be extended to the origin, so that we can take the cone to be 
$\R^9$. Thus
\bea
{\rm Parallel~spinors~on~} \R^9 &\Longleftrightarrow &
{\rm Real~Killing~spinors~on~} S^8
\nonumber\\
\nabla_{\mu} \hat{\epsilon}=0  &\longleftrightarrow &
(\nabla_{\mu} -\frac{1}{2} \gamma_{\mu})\epsilon =0 \; .
\eea
The observation in \cite{JFO} is that this decomposition, written as
$\frak{l} =\frak{l}_0 \oplus \frak{l}_1$, has the 
interpretation in terms of Killing superalgebras on $S^8$:
$\frak{l}_0=\frak{so}(9)$ is the Lie algebra of isometries of $S^8$ and 
$\frak{l}_1= S^+$ is the space of Killing spinors on $S^8$. The latter 
comes, via the cone construction, from real Killing spinors on 
the cone $\R^9$. Hence
\(
\frak{f}_4=\left\{ {\rm Even~isometries~on~} S^8 \right\}
\oplus  \left\{ {\rm Odd~isometries~on~} S^8 \right\},
\)  
and the Lie brackets for the super Lie algebra are satisfied \cite{JFO}.
Schematically (abusing notation of fiber vs. bundle), we have
\(
\xymatrix{
{\underbrace{ {\rm Spin}(9)-{ \rm structures} }_{{\mbox V_9}} }
\ar[ddrr]
&&
~~~~{\underbrace{{\rm Killing~ spinors}}_{{\mbox S^8}} }~~~
\ar[dd]
\ar@{_{(}->}[ll]
\ar@{^{(}->}[rr]
&&
~~{\underbrace{ {\rm parallel~spinors} }_{{ {\mathcal C} S^8}} }
\ar[ddll]
\\
\\
&&
{\bb O} P^2
&&
}
\)
From this and the earlier discussion we therefore have

\begin{prop}
$\frak{f}_4$ is the Lie superalgebra of a sphere inside $V^9$. Hence 
the unification of the fields in M-theory and their supersymmetry can 
be seen from the eight-sphere over ${\bb O} P^2$. 
\label{prop:f4}
\end{prop}
 
 We can give another interpretation to the Euler triplet in terms of spinors.
 We have seen in the Remark containing equation (\ref{twisted index}) 
 that the Euler triplet can be interpreted as an index of a twisted Dirac
 operator. The kernel of the operator  
 (\ref{def twi dir}) is the space of harmonic spinors, which is the
 desired representation up to sign \cite{Land1}. Therefore, we get another 
 characterization of the supergravity multiplet.
 
 \begin{prop}
 The identity representation of $F_4$ encoding the supergravity multiplet is 
 the space of twisted harmonic spinors on $\mathbb{O} P^2$. 
\label{thm:harmonic}
 \end{prop}

 \vspace{3mm}
\noindent {\bf Comparison to generation of supersymmetry from lattices.}
 Next we discuss the relation, similarities and differences between the
 above process of generating fermions and supersymmetry and the 
one through which the various closed superstring theories are derived 
starting from
the closed bosonic string \cite{CENT}. The spectrum of the bosonic string contains no
fermions and so these are generated on a lattice in internal space. 
 In \cite{CENT} the following procedure was created:

\noindent (1) Seek an internal symmetry group $G$ containing the little group ${\rm Spin}(8)$.
This is achieved by a torus compactification $T/\Lambda_G$, with $\Lambda_G$
the root lattice of a simply-laced group $G$ of rank $8$.

\noindent (2) Declare the diagonal subgroup $SO(8)_{\rm diag} \subset SO(8) \times {\rm Spin}(8)$
as the new transverse group. This implies that the spinor representations of
${\rm Spin}(8)$ describe fermionic states. 

\noindent (3) Extend $SO(8)_{\rm diag}$ to the full Lorentz group $SO(1, 9)_{\rm diag}$.

\noindent (4) Impose the supersymmetry requirement that a consistent truncation on
the spectrum of the bosonic string be performed. This requires a regular 
embedding so that the root lattice $\Lambda_{{\rm Spin}(8)}$ is contained 
in $\Lambda_G$.

\vspace{3mm}
The only simply-laced groups which contain ${\rm Spin}(8)$ as a subgroup
in a regular embedding are $E_6$, $E_7$ and $E_8$. Requiring the rank to
be $8$ then singles out $G=E_8 \times E_8$. Then \cite{CENT}: 

\noindent (i) the choice $G_L=G_R=E_8 \times E_8$
for the groups in the left and right sector gives the two type II string theories;

\noindent (ii) the same choice with a truncation on the 
left-moving sector gives the $E_8 \times E_8$ heterotic string;

\noindent (iii) the 
choice $G_L=E_8 \times E_8$, $G_R={\rm Spin}(32)/\Z_2$ together with
a truncation on the left-moving sector gives the ${\rm Spin}(32)/\Z_2$
heterotic string theory.

\vspace{3mm}
Now let us compare the similarities and the differences of our case with 
the above formalism of \cite{CENT}. We record this in the following remarks.

\vspace{3mm}
\noindent {\bf Remarks}

\noindent{\bf 1.} The M-theory  case is geometric and involves nontrivial topology.
This is in contrast to the torus in a vertex-operator-like construction in the string case.

\noindent {\bf 2.} $F_4$ is not simply-laced and hence cannot be involved in the
internal torus construction. 

\noindent {\bf 3.} In both cases, the fermions are generated from the internal space.
However, in \cite{CENT}, fermionic states are generated from bosonic states.
In fact, in our case, the whole massless spectrum of 
 eleven-dimensional supergravity is generated from the two Spin bundles 
 in dimension sixteen. This method of generating fermions 
 is very different from the string formalism of generating fermions from 
torus compactification. 

\noindent {\bf 4.} The signature $\sigma(M^{4k})$ of an oriented
 $4k$-dimensional manifold $M^{4k}$ is an invariant of the manifold. 
 Moreover, the signature of $-M^{4k}$, which is $M^{4k}$ with the 
 orientation reversed, is equal to the negative of the signature of 
 $M^{4k}$: $\sigma(-M^{4k}) = - \sigma (M^{4k})$. Since 
 $\sigma({\bb O} P^2) \neq 0$, this means that there
 is no orientation-reversing homeomorphism $f : {\bb O} P^2 \to {\bb O} P^2$ such that
 $f_* [{\bb O} P^2] = - [{\bb O} P^2]$. The implication is, in particular, 
 that we cannot impose
 any such involution on the fermions.

\noindent {\bf 5.} The construction in M-theory using $F_4$ involves the 
Spin bundle of ${\bb O} P^2$. This means that in twenty-seven dimensions
the theory will have fermions. This is a major difference from the bosonic
string case, which has no fermions in its spectrum. How can this be 
compatible with the bosonic string and with the classification of supersymmetry 
in general? In relation to the bosonic string, it could be that there is an
involution that kills the fermions in a way similar to what happens to 
the $C$-field in going from M-theory to the heterotic string, or from 
the conjectural bosonic M-theory in \cite{HS} to bosonic string theory.
Let us now consider the second part of the question related 
to the classification of supersymmetry. The action
in twenty-seven dimensions might involve fermions, and so the question
is whether this will/can be supersymmetric. 
That is something to be investigated. However,
for now we can say that being supersymmetric does not contradict the no-go
theorems in supersymmetry as those involve the Lorentz condition. 
The sixteen-dimensional internal space can be taken to have either 
all time or all space signature, i.e. $(16, 0)$ or $(0, 16)$, respectively. 
We then get for the signature $(t, s)$ of the 27-dimensional space
\bea
(1, 10) + (0, 16) &=& (1, 26)
\\
(1, 10) + (16, 0) &=& (17, 10).
\eea
The first one obviously wildly violates the no-go theorems but the 
second does not as $t-s=7$. Note that a version of eleven-dimensional
M-theory with $s-t=7$ was constructed by Hull \cite{Hull}. 
While supersymmetry seems 
mathematically admissible, it is far from obvious what to make
physically of so many such time directions. We do not address this here. 

\subsection{Relating $Y^{11}$ and $M^{27}$}  

\subsubsection{geometric consequences}
\label{geom conseq}
We start with the Riemannian geometry of ${\bb O} P^2$. 
Consider the following three subsets of ${\bb O}^3$
\(
U_1=\{ 1\} \times {\bb O} \times {\bb O} , ~~~~~~~~
U_2= {\bb O} \times \{ 1\} \times {\bb O} , ~~~~~~~~
U_3= {\bb O} \times {\bb O} \times \{1\},  
\)
and form the union $\mathcal{U}:=U_1 \cup U_2 \cup U_3$.
Define the following relation $\sim$ on ${\bb O}^3$:
\(
[a, b, c] \sim [d, e, f] \longleftrightarrow {\rm ~ there ~exists~}
\lambda \in {\bb O} - \{ 0\} {\rm~~ such~that~} a=d\lambda, b=e \lambda, 
c=f\lambda.
\)
The relation $\sim$ on $\mathcal{U}$ is an equivalence relation
\cite{As}. 
The Cayley projective plane is the set of equivalence classes of
$\mathcal{U}$ by the equivalence relation $\sim$, that is
$
{\bb O} P^2=\mathcal{U}/ \sim$.
Keeping in mind ${\bb O} \cong \R^8$, an atlas on ${\bb O} P^2$ can be 
taken to be $(U_i/ \sim, \phi_i)$,
$i=1,2,3$, where the homeomorphisms $\phi_i$ are given by
\bea
\phi_1 &:& U_1/\sim ~ \longrightarrow \R^{16}, ~~~~~~~~ \phi_1([a, b, c])=(b, c),
\nonumber\\
\phi_2 &:& U_2/\sim ~ \longrightarrow \R^{16}, ~~~~~~~~ \phi_2([a, b, c])=(a, c),
\nonumber\\
\phi_3 &:& U_3/\sim ~ \longrightarrow \R^{16}, ~~~~~~~~ \phi_3([a, b, c])=(a, b)\;.
\eea
The transition functions $\phi_i \circ \phi_j^{-1}: \R^{16} \to \R^{16}$
\bea
\phi_1 \circ \phi_2^{-1} (a, b) &= & (a^{-1}, ba^{-1})=
\phi_2 \circ \phi_1^{-1} (a, b),
\nonumber\\
\phi_1 \circ \phi_3^{-1} (a, b) &= &(ba^{-1}, a^{-1})=
\phi_3 \circ \phi_1^{-1} (a, b),
\nonumber\\
\phi_2 \circ \phi_3^{-1} (a, b) & =& (b^{-1}, ab^{-1})=
\phi_3 \circ \phi_2^{-1} (a, b)
\eea
are diffeomorphisms and hence we get a smooth
$16$-dimensional manifold structure for ${\bb O} P^2$
\cite{HSV}.

\vspace{3mm}
The metric on ${\bb O} P^2$ can be obtained from the metrics on 
the charts which are compatible with respect to transition maps. 
The metric, with $(u, v)$ coordinate functions, is \cite{HSV}
\(
ds^2 = \frac{ |du|^2 (1+ |v|^2) + |dv|^2(1+|u|^2) - 2 {\rm Re}
[(u \overline{v}) (dv d{\overline{u}})]}
{(1+|u|^2 + |v|^2)^2}.
\)
In terms of a coordinate frame $\{e_1, \cdots, e_8, f_1, \cdots, f_8 \}$
where $e_i = \partial_i$ and $f_i=\partial_{i+8}$ for $1 \leq i \leq 8$, 
the unmixed components of the metric are
\(
g(e_i , e_j)= \delta_{ij} \frac{1 + |v|^2}{(1+|u|^2 + |v|^2)^2}\;,
\qquad \quad
g(f_i , f_j)= \delta_{ij} \frac{1 + |u|^2}{(1+|u|^2 + |v|^2)^2}\;.
\)
The mixed components, in terms of the standard orthonormal 
basis $\{ x_1, \cdots, x_8\}$ of ${\bb O}$ are 
\(
g(e_i, f_j)=g(f_i, e_j)= - \frac{\langle (u \overline{v}) x_j, x_i\rangle}
{(1+ |u|^2 + |v|^2)^2}. 
\)
Using the identity
$
R_{\mu \nu \lambda \sigma}=R_{\mu \nu \lambda}{}^{\sigma} =
\Gamma_{\mu \lambda; \nu }^{\sigma}
-
\Gamma_{\nu \lambda; \mu }^{\sigma}
= \frac{1}{2} \left[ 
g_{\nu \lambda ; \mu \sigma} 
+ g_{\mu \sigma ; \nu \lambda}
- g_{\mu \lambda ; \nu \sigma}
- g_{\nu \sigma ; \mu \lambda}
\right]$,
the only non-vanishing components of the Riemann tensor are 
\cite{HSV}
\bea
&&R(e_i, e_j, e_i, e_j) = -R( e_i, e_j, e_j, e_i)=4,
\nonumber\\
&&R(f_i, f_j, f_i, f_j) = -R(f_i, f_j, f_i, f_j)=4,
\nonumber\\
&&R(e_i, e_j, f_k, f_l) = R(f_k, f_l, e_i, e_j)= - \langle x_i \overline{x}_l, 
x_j \overline{x}_k \rangle +
\langle x_j \overline{x}_l, 
x_i \overline{x}_k \rangle ,
\nonumber\\
&&R(e_i, f_j, e_k. f_l) = R(f_i, e_j, f_k, e_l)=
\langle x_i \overline{x}_j, 
x_k \overline{x}_l \rangle,
\nonumber\\
&&R(f_i, e_j, e_l, f_k) = - \langle x_i \overline{x}_j, 
x_k \overline{x}_l \rangle \; .
\eea
It can now be easily seen that both the Ricci curvature tensor $R_{\mu \nu}$ 
and the Ricci scalar $R$ are both positive.

\vspace{3mm}
Taking $M^{27}$ to be the total space of an ${\bb O} P^2$ bundle over
$Y^{11}$ then the Ricci curvatures of the two spaces are related. In
particular, since ${\bb O} P^2$ is a compact Riemannian manifold which 
has a metric of positive Ricci curvature on which the Lie group $F_4$ 
acts by isometries, and the base $Y^{11}$ is a compact manifold, 
it follows from O'Neill's formulae for submersions (see \cite{Besse}) that
 
\begin{prop}
If the base $Y^{11}$ admits a metric of positive Ricci curvature, then
so does the 27-dimensional space.
\end{prop}

This is shown by taking a certain metric on $M^{27}$ with totally geodesic 
fibers (\cite{Besse}) and then shrinking the ${\bb O} P^2$
fibers $\grave{\rm a}$ la Kaluza-Klein. This is a specific case of the ${\bb O} P^2$ analog 
of Proposition 3.6 in \cite{Conj}.

\subsection{Structures on $M^{27}$}
\label{strs}

\vspace{3mm}
The cohomology of ${\bb O} P^2$ is $H^*({\bb O} P^2; C)=C[x]/x^3$, 
$|x|={\rm deg}\hspace{0.5mm}x=8$,
as an algebra.

\vspace{3mm}
\noindent {\bf Remarks}

\noindent {\bf 1.} Note that a priori the characteristic of $C$ should 
divide the order of the Weyl group of $F_4$. Since $|W(F_4)|=2^7 \cdot 3^2$
then the candidate primes are 2 and 3 only. We have seen that among
these two numbers only the prime 3 gives a nontrivial Serre fibration.

\noindent {\bf 2.} Note that the primes 2 and 3 are also the torsion primes
of $F_4$. It is not the case in general that the torsion primes for $G$
are exactly the same primes that appear in the factorization of $|W(G)|$.

\vspace{3mm}
The total space of an $\mathbb{H}P^2$ bundle over a Spin manifold is
again a Spin manifold. However, the same property is not
automatically true for total spaces of ${\bb O} P^2$ bundles over $BO
\langle 8 \rangle$-manifolds. The reason is that while the tangent
bundle $T$ along the fibers of the universal bundle 
\( 
{\bb O} P^2=F_4/{\rm Spin}(9) \longrightarrow
B{\rm Spin}(9) \longrightarrow BF_4 
\label{main} 
\) 
 has
a Spin structure --- since $H^i(B{\rm Spin}(9))=0$ for $i=1,2,3$ --- it
has no $BO\langle 8 \rangle$ structure. This can be explained as follows,
using \cite{Klaus}. The complementary roots of $i : {\rm Spin}(9)
\hookrightarrow F_4$ are the 16 roots $\frac{1}{2}(\pm x_1 \pm x_2
\pm x_3 \pm x_4)$, where $x_i$ denote the standard linear forms on
$\frak{so}(9)$. Using Borel-Hirzebruch methods \cite{BH}, the total Pontrjagin
class $p(T) \in H^*(B{\rm Spin}(9);\Q)$ is given by the product
$\frac{1}{4}\prod (\pm x_1 \pm x_2 \pm x_3 \pm x_4)$, so that the
first Pontrjagin class is \( p_1(T)=2(x_1^2 + x_2^2 + x_3^2 + x_4^2)
\in H^4(B{\rm Spin}(9);\Q).\) This is of course invariant under the
Weyl group of ${\rm Spin}(9)$. However, it is also invariant under
$W(F_4)$, and hence belongs to $H^4(BF_4;\Q)=\Q$ as well. This shows
that $p_1(T)$ can be considered as coming from the universal space
for ${\rm Spin}(9)$ or $F_4$.

\begin{prop}
If $Y^{11}$ admits a String structure then so does $M^{27}$ provided that
there is no contribution from the degree four class from $BF_4$. 
\label{cond}
\end{prop}

\proof
We have the ${\bb O} P^2$ bundle over $Y^{11}$ with total space $M^{27}$
\( 
\xymatrix{M^{27} \ar[r]^{\tilde{f}} \ar[d]_{\pi} & B{\rm Spin}(9)
\ar[d]^{Bi}\\ Y^{11} \ar[r]_{f} & BF_4}, 
\) 
which gives the
decomposition $TM^{27}=\pi^*TY^{11}\oplus {\tilde f}^*T$, and so the
tangential Pontrjagin class is 
\( 
p_1(M^{27})= \pi^*\left(
p_1(Y^{11})+f^*p_1(T) \right). 
\) 
In the case $Y^{11}$ is a
3-connected $BO\langle 8 \rangle$-manifold, we have that
$H^4(Y^{11};\Z)$ is free and $\pi^* : H^4(Y^{11};\Z) \to
H^4(M^{27};\Z)$ is an isomorphism.
Thus $M^{27}$ is also a $BO\langle 8 \rangle$-manifold if and only
if $f^*\overline{x}_4=0 \in H^4(Y^{11};\Z)$, where $\overline{x}_4
\in H^4(BF_4;\Z)$ is the generator. Therefore we have shown that $M^{27}$
is String if and only if $G_4$ in M-theory gets no contribution from
$BF_4$.
\endproof

\noindent {\bf Remarks}

\noindent {\bf 1.} The quantization condition for the field strength $G_4$ in
M-theory is known \cite{Flux}. Since this field does not seem to get a contribution 
from a class in $BF_4$, the condition in Proposition \ref{cond} seems
reasonable. In some sense we could view the presence of such a degree 
four class as an anomaly which we have just cured. Alternatively, 
one can discover that this is not as serious as it might seem--- see
the more complete discussion in section \ref{effect on PF}.

\noindent {\bf 2.} We connect the above discussion back to cobordism
groups. While there is no transfer map from $\Omega_{11}^{\langle 8
\rangle}(BF_4)$ to $\Omega_{27}^{\langle 8 \rangle}$, there is a
transfer map after killing $\overline{x}_4$ \cite{Klaus}. Denoting
by \footnote{This is the analog of the String group when $G={\rm
Spin}$, in the sense that it is the 3-connected cover.} $BF_4\langle
\overline{x}_4\rangle$ the corresponding classifying space that
fibers over $BF_4$, killing $\overline{x}_4$ is done by pulling back
the path fibration $PK(\Z,4) \to K(\Z, 4)$ with a map $\overline{x}_4
: BF_4 \to K(\Z, 4)$ realizing $\overline{x}_4$. The corresponding
transfer map is $\Omega_{11}^{\langle 8 \rangle}(BF_4\langle
\overline{x}_4 \rangle) \to \Omega_{27}^{\langle 8 \rangle}$.

\vspace{3mm}
Next, for the higher structures we consider String and Fivebrane structures
\cite{SSS1} \cite{SSS2}. We have

\begin{prop}
\noindent {\bf 1.} In order for $M^{27}$ to admit a Fivebrane structure, 
the second Pontrjagin class of $Y^{11}$ should be the negative of that
of ${\bb O} P^2$, i.e. $p_2(TY^{11})= -p_2(T {\bb O} P^2)=- 6u$. 

\noindent {\bf 2.} $\widehat{A} (M^{27})=0$, irrespective of whether 
or not the $\widehat{A}$-genus of $Y^{11}$ is zero.

\noindent {\bf 3.} The Witten genus $\Phi_W(M^{27})=0$.

\noindent {\bf 4.} The elliptic genus $\Phi_{\rm ell}(M^{27})=0$.
\label{ellm27}
\end{prop}
\proof
For part {\bf 1} note that if $Y^{11}$ admits a Fivebrane structure 
then $M^{27}$ does not necessarily admit such a structure.
This is because the obstruction to having a Fivebrane structure is
$\frac{1}{6}p_2$ \cite{SSS2} but 
we know that $\frac{1}{6}p_2({\bb O} P^2)=u \neq 0$. 
However, we can choose $Y^{11}$ appropriately so that it 
conspires with ${\bb O} P^2$ to cancel the obstruction and lead
to a Fivebrane structure for $M^{27}$. 
Noting that the tangent bundles are related as $TM^{27}= TY^{11} \oplus T{\bb O} P^2$,
then considering the degree eight
part of the formula (see \cite{Mil})
\(
p(E \oplus F)= \sum p(E) p(F)~~~~{\rm mod}~2{\rm -torsion}.
\label{ps}
\)
we get for our spaces
\(
p_2(TY^{11} \oplus T {\bb O} P^2)= p_1(TY^{11}) p_1(T {\bb O} P^2)
+ p_2(TY^{11}) + p_2(T {\bb O} P^2)~~~{\rm mod~}2{\rm -torsion}.
\)
Since we have $p_1(T {\bb O} P^2)=0$, then requiring that
$p_2(TM^{27})=0$ leads to the constraint that
$p_2(TY^{11})+ p_2(T {\bb O} P^2)=0$ modulo 2-torsion.

For part {\bf 2} we use the multiplicative property of the $\widehat{A}$-genus
for Spin fiber bundles to get
$
\widehat{A}(M^{27})= \widehat{A} (Y^{11}) 
\widehat{A} ({\bb O} P^2)$.
Since the $\widehat{A}$-genus of ${\bb O} P^2$ is zero then the result follows.

For part {\bf 3} we use a result of Ochanine \cite{Och1}. Taking the total space 
$M^{27}$ and the base $Y^{11}$ to be closed oriented manifolds, and since 
the fiber ${\bb O} P^2$ is a Spin manifold and the structure group $F_4$ 
of the bundle is compact, then the multiplicative property of the genus can be 
applied
\(
\Phi_W (M^{27})= \Phi_W ({\bb O} P^2) \Phi_W (Y^{11}).
\)
Now since we proved in \cite{OP2} that $\Phi_W ({\bb O} P^2)=0$, it 
follows immediately that $\Phi (M^{27})$ is zero regardless of whether
or not $\Phi_W (Y^{11})$ vanishes. Even more, $\Phi_W(Y^{11})$ is zero
because $Y^{11}$ is odd-dimensional. 
\footnote{ However, see the case when
$Y^{11}$ is a circle bundle at the end of this section.} 

For part {\bf 4} we use the fact that the fiber is Spin and the structure group
$F_4$ is compact and connected so we can apply the multiplicative 
property of the elliptic genus \cite{Och1}
\(
\Phi_{\rm ell}(M^{27})= \Phi_{\rm ell}(Y^{11}) \Phi_{\rm ell}({\bb O} P^2).
\)
In this case the genus for the fiber is not zero (see \cite{OP2})
but the elliptic genus of $Y^{11}$ is zero, again because of dimension.
Therefore $\Phi_{\rm ell}(M^{27})=0$. 
\endproof

\paragraph{Ochanine genera.}

\vspace{3mm}
There is another description of the Ochanine $k$-invariant \cite{Klaus2},
which we will use to make a connection
to invariants appearing in M-theory.

\begin{prop}
1. The Ochanine invariant of a ten-dimensional closed Spin manifold
$X^{10}$ is equal to the mod 2 index of the Dirac operator twisted with
the virtual bundle $TX^{10}- 2$. 
\label{thm:mod2}
\end{prop}
\proof
The family index theorem says that 
for $E$ a real bundle in $KO^0(X^{10})$ an invariant $e\in \Z_2$ was 
defined by Atiyah and Singer \cite{AS} by 
$\langle E , [X^{10}]_{KO} \rangle=e \eta^2 \mu \in KO_{10}$,
which turns out to be the mod 2 index of the Dirac operator
$D_E$ of $X^{10}$ twisted by the virtual bundle $E$,
\(
e={\rm dim}_{\C}\hspace{0.5mm}{\rm ker}(D_E)~~~{\rm mod}~2.
\)
Applying \cite{Klaus}, the $k$-invariant of $X^{10}$ is the 
coefficient of $q$ in the expression
$f(q)^{-8} \Phi_{\rm och} \in KO_{10}[[q]]$, where 
\(
f(q):= \sum_{n \geq 1} q^{\binom{n}{2}}=
1+ q + q^3 + q^6 + \cdots,
\)
since $\varepsilon /q=f(q^8)$ mod $2=f(q)^8$ mod 2. 
We find the coefficient of $q$ in the expansions. We have
\(
f(q)^{-8}=(1+ q + \cdots)^{-8}= 1 -8q + \cdots. 
\label{fq -8}
\)
The expansion for $\theta (q)$ takes the form
\(
\theta (q)=\left(\frac{1-q}{1-q^2} \right)
\left(\frac{1-q^3}{1-q^4} \right) \cdots = 1- q + \cdots,
\label{theta q}
\)
so that $\theta (q)^{-10}=1 + 10q + \cdots$. The expansion of $\Theta_q(E)$ is
\bea
\Theta_q (E) &=& \Lambda_{-q}(E) \otimes S_{q^2} (E) + \cdots
\nonumber\\
&=& \left( \sum_{k \geq 0} {(-q)}^k {\Lambda}^k (E) \right) \otimes
\left( \sum_{k \geq 0} (q^2)^k S^k (E) \right)
\nonumber\\
&=& 1 -qE + \cdots .
\label{Theta q}
\eea
Putting the expressions (\ref{fq -8}), (\ref{theta q}), (\ref{Theta q})
together we get
\bea
f(q)^{-8} \theta (q)^{-10} \Theta_q (TX^{10} )
&=& (1-8q+ \cdots ) (1+ 10q + \cdots) (1 - q TX^{10}+ \cdots )
\nonumber\\
&=& 1+ (2- TX^{10})q + \cdots .
\eea
Extracting the coefficient of $q$ we get the desired result. 

Note that there is another way of obtaining this which makes use
of the grading for $\Phi_{\rm och}$. Instead of looking at $\theta_q$ and
$\Theta_q$ separately, we can look at the coefficient of $q$ in 
the Ochanine genus $\Phi_{\rm och}(X^{10})$.
This is
\(
\Phi_{\rm och}^1 = \langle - \Pi_1 (TX^{10}), [X^{10}]_{KO}\rangle \in KO_{10}=\Z_2,
\label{phi1}
\)
where $\Pi_1$ is the first KO-Pontrjagin class (defined in \cite{ABP2}), 
which is equal to 
$\Lambda^1(TX^{10} -10)=TX^{10} -10$. Substituting in (\ref{phi1})
we get
\(
\Phi_{\rm och}^1 = \langle -(TX^{10}-10)\; , \; [X^{10}]_{KO}\rangle, 
\)
which agrees with the product $\theta_q^{-10} \Theta_q (TX^{10})= 1 + (10 -TX^{10})q + \cdots$.
\endproof

\noindent {\bf Remarks}

\noindent {\bf 1.} Note that, interestingly, the bundle we get is the Rarita-Schwinger bundle
with the dilatino and the spinor ghosts, as the Rarita-Schwinger field $\Psi$
which leads to gauge invariance is 
a section of $SX^{10} \otimes (TX^{10} - 2\mathcal{O})$, where $\mathcal{O}$ 
is the trivial line bundle. The (mod 2) index $I_{RS}$ of the corresponding Dirac operator 
$D_{RS}$ appears in the phase of the partition function \cite{DMW} 
through the phase of the Pfaffian
\(
Pf(D_{RS})=(-1)^{I_{RS}/2}~ |Pf(D_{RS}|.
\)
What is remarkable is that the `quantum' Rarita-Schwinger operator appears 
directly in this formulation.
 
\noindent {\bf 2.} In \cite{DMW} the main focus was the dependence of the partition function
on the degree four class $a$ coming from the $E_8$ gauge theory, but 
the contribution from $I_{RS}$ was also given.  
The main example discussed in \cite{DMW} is $X^{10}=\H P^2 \times T^2$. 
Using the property
\(
k(M^8 \times S^1 \times S^1) =\sigma(M^8)~~{\rm mod}~2\;,
\label{eight}
\)
we can indeed see that the Ochanine $k$-invariant in this case is not zero.
With $T^2$ taken as the product of two circles with nontrivial Spin structures
we have
\(
k (S^1 \times S^1 \times \H P^2)= \sigma (\H P^2)~~{\rm mod}~2,
\)
which is equal to 1, since $\sigma(\H P^2)=1$.

\noindent {\bf 3.} In defining the elliptically refined partition function in M-theory 
and type IIA string theory, a real-oriented elliptic cohomology theory appears
\cite{KS1}. This is $EO(2)$, the fixed point, with respect to the formal inverse,
of the theory $E\R(2)$, the real version of  Morava theory $E(2)$, which 
has two generators $v_1$ and $v_2$. The orientation in this theory is shown to
be given by $w_4$ \cite{KS1}. It was also shown that when $w_4(X^{10})=0$, 
$X^{10}$ has an
$EO(2)$-orientation class $[X^{10}]_{EO(2)_{10}} \in EO(2)_{10}(X^{10})$,
and for $x \in E^0(X^{10})$, the refined mod 2 index in this theory is
\(
j(x)=\langle x~ \overline{x}\; ,\; [X^{10}]_{EO(2)}\rangle
\in EO(2)_{10}=\Z_2[v_1^3v_2^{-1}]. 
\)

\section{Terms in the Lifted Action}
\label{sec:terms}

In this section we consider possible terms in the lifted
action up in twenty-seven dimensions. We first consider 
geometric expressions involving a distinguished 8-form,
called the Cayley 
8-form, in section  \ref{sec:cayley}. Then in section \ref{effect on PF}
we consider torsion classes and their effect on the M-theory
partition function.

\subsection{The Cayley 8-form}
\label{sec:cayley}
In section \ref{susy} we discussed
the question of whether the higher-dimensional `theory'
in our case is supersymmetric.  In any case holonomy
would give us a handle on whatever differential forms end up
appearing. The holonomy group of ${\bb O} P^2$ is ${\rm Spin}(9)$
and there is in fact a ${\rm Spin}(9)$-invariant 
8-form that generalizes the K\"ahler 2-form for $\CP^2$ and the
fundamental or Cayley 4-form on $\H P^2$ \cite{BG}. 
The ${\rm Spin}(9)$ representation $\Lambda^8 (\Delta_9)= 
\Lambda^8 (\R^{16})$ contains a unique 8-form which is
invariant under the action of ${\rm Spin}(9)$. 
Note that ${\bb O} P^2$ does not admit an almost complex structure
\cite{BH} nor an almost quaternionic structure \cite{Bar2}. 

\vspace{3mm}
There is an explicit
expression for the 8-form  
in 
the tangent plane ${\bb O} \oplus {\bb O}$
to ${\bb O} P^2$. An original one in terms of cross-product was given in  
\cite{BG}, and other versions later in \cite{BP} \cite{AM} and \cite{CGM}. 
\footnote{We thank the authors of \cite{CGM} for pointing out that
expressions prior to theirs have problems.}
We will use the latter.
A $\mathrm{Spin}(9)$-structure is given by \cite{Fried} 
a nine-dimensional subbundle
of the bundle of endomorphisms $\mathrm{End}(TM)$ locally
spanned by nine orthogonal linear transformations
$I_i$, $0 \leqslant i \leqslant 8$, of $\R^{16}$
satisfying the relations $I_i I_j + I_j I_i = 0$, $i \neq j$,
$I^2_i = \mathrm{I}$, $I^T_i = I_i$, $\mathrm{tr}\; I_i = 0$,
$i,j=0,\dots,8$. These endomorphisms define $2$-forms
$J_{ij}$, $0\leqslant i < j\leqslant 8$, on $M$ locally by
$J_{ij}(X,Y) = g(X,I_i I_j Y)$. Similarly, using the
skew-symmetric involutions $I_i I_j I_k$, $0\leqslant
i<j<k\leqslant 8$, one can define $2$-forms $\sigma_{ijk}$. The
$2$-forms $\{ J_{ij}, \sigma_{ijk} \}$ are linearly
independent and a local basis of the bundle $\Lambda^2M$.
Then the canonical 8-form is given by \cite{CGM}
\(
\omega_8= \sum_{\substack{
i,j=0,\dots,8\\ i',j'=0,\dots,8}} J_{ij} \wedge
J_{ij'} \wedge J_{i'j} \wedge J_{i'j'}\;,
\)
where
$J_{ij}=-J_{ji}$ if $i>j$ and $J_{ij}=0$
if $i=j$.
Note that the 8-form has the following properties:

\noindent (1) The 8-form defines a unique parallel form on ${\bb O} P^2$.

\noindent (2) Since the signature of ${\bb O} P^2$ is positive, then 
the 8-form is self-dual. 

\noindent (3) The 8-form is the 8-dimensional analog of the K\"ahler
2-form and the quaternion-K\"ahler 4-form. 

\vspace{3mm}
\noindent {\bf Remarks}

\noindent {\bf 1.} At the rational level we can thus use $\omega_8$ to build 
a ${\rm Spin}(9)$-invariant degree sixteen expression  
$
\rho_{16}^{\R}= \omega_8 \wedge \omega_8$
that we integrate and insert as part of the action as $\int_{{\bb O} P^2} \rho_{16}^{\R}$. 

\noindent {\bf 2.} Assume that there are 
fields $\mathcal{F}_8$ and $\mathcal{F}_{16}$ in the 27-dimensional 
`theory' with potentials ${\mathcal C}_7$ and ${\mathcal C}_{15}$. 
 In the dimensional reduction on ${\bb O} P^2$ to eleven dimensions,
a natural ${\rm Spin}(9)$-invariant 
ansatz for the fields may be taken, at the rational level, to be 
\(
{\mathcal F}_8 = \omega_8, ~~~~~~~~~~~~~~~~~~ 
{\mathcal F}_{16} =\omega_8 \wedge \omega_8\;,
\label{cal F}
\)
and similar expressions at the integral level 
Note that
since $\omega_{16}$ is essentially the volume form, then such an ansatz is 
the analog of the Freund-Rubin ansatz \cite{FR} in the 
reduction of eleven-dimensional supergravity to lower dimensions.

\subsection{Torsion classes and effect on the M-theory partition function}
\label{effect on PF}
In subtle situations the fields in the physical theory 
can be torsion classes in cohomology. 
We consider terms in the action coming from $BF_4$ or from the 
fiber ${\bb O} P^2$.  We will show that torsion classes from $BF_4$
are compatible with the description in \cite{DMW} of the phase of the
M-theory partition function.

\subsubsection{Classes from $BF_4$}

\noindent {\bf 1. $\Z_2$ coefficents:}
The cohomology ring of $BF_4$ with coefficients in $\Z_2$ is given from
 by the polynomial ring \cite{Borel} 
\( 
H^*( BF_4;\Z_2)=\Z_2 \left[ x_4, Sq^2 x_4,
Sq^3 x_4, x_{16}, Sq^8 x_{16}\right], 
\label{bfz2}
\) 
where $x_4$ and $x_{16}$
are polynomial generators of degree 4 and 16, respectively.
From the structure of the cohomology ring (\ref{bfz2}) we see that
we can pull back classes from $BF_4$ and that these are in fact
compatible 
with the fields of M-theory. In particular, there is a degree four class
$x_4$, as in all Lie groups of dimension greater than or equal to
three, which could be matched with the field strength $G_4$ in M-theory.
In fact, since {\it any} degree four class can be the characteristic class
$a_{E_8}$ of an $E_8$ bundle, then a class pulled back from $F_4$
can certainly be at the same time a class of some $E_8$ bundle. 
Hence an $F_4$ class is possible in the shifted quantization 
condition
$
[G_4] - \frac{\lambda}{2}= a_{E_8} \in H^4(Y^{11}; \Z)$, 
discovered in \cite{Flux}.

\vspace{3mm}
The higher degree classes are also relevant. We also have
the degree six and the seven generators $Sq^2 x_4$ and $Sq^3 x_4$, respectively, 
which, when nonzero, would appear in the
phase of the partition function. The comparison of M-theory on $Y^{11}$
 with type 
IIA string theory on a ten-manifold $X^{10}$ involves the bilinear form  
\cite{DMW}
$
\mu (a, b) = \int_{X^{10}} a \cup Sq^2 b
$,
for $a, b \in H^4(X^{10}; \Z)$. This can be viewed \cite{DMW}  as 
a torsion pairing 
\bea
T : H^4_{\rm tor}(X^{10}; \Z) \times  H^7_{\rm tor}(X^{10}; \Z)
&\longrightarrow& U(1)
\nonumber\\
\left( \; a \;, \; Sq^3 b\; \right) &\longmapsto& \int_{X^{10}} a \cup Sq^2 c,
\eea
where $Sq^3 b=\beta(Sq^2 c)=Sq^1 Sq^2 c= Sq^3 c$. In our case
$a$ and $b$ can be $f^* x_4$. Thus we have
\begin{prop}
$\Z_2$ classes from $BF_4$ are compatible with the M-theory 
partition function, i.e. they produce no anomalies and they do not
change the value of the partition function.
\label{thm:comp}
\end{prop}

\noindent {\bf 2. $\Z_3$ coefficents:}
If we restrict to low degrees, say $\leq 16$, then we have the truncated polynomial
\(
H^*(BF_4; \Z_3) \cong  \Z_3[x_4, x_8] + \Lambda(x_9).
\label{trun}
\)
Now the main observation is that the class $x_9$, being 
$\beta P_3^1 x_4$, is the same as the class required to be cancelled
in theorem \ref{d9thm}. If we kill this class then we are left
with only the degree four and the degree eight classes $x_4$ and
$x_8$. Since $x_8$ is $P_3^1 x_4$, then this $\Z_3[x_4, P_3^1x_4]$ 
is also compatible with
the mod 3 description of the anomalies in M-theory described in
\cite{KSpin}. Therefore,

\begin{prop}
$\Z_3$ classes from $BF_4$ are compatible with the partition 
function of M-theory once the anomaly $P_3^1 x_4$ is 
cancelled.
\label{thm:cancel}
\end{prop}

\subsubsection{Classes from ${\bb O} P^2$}

Recall that we have introduced fields ${\mathcal F}_8$ and 
${\mathcal F}_{16}$ with corresponding potentials 
${\mathcal C}_7$ and ${\mathcal C}_{15}$, respectively 
(see (\ref{cal F})). Assuming that the 27-dimensional 
`theory' indeed has such fields, we consider some consequences
in this section.  We emphasize that 
we do not have enough knowledge about the dynamics (if and when it exists) 
in 27 dimensions so we will concentrate on the topology. We will 
concentrate on the first field, because of the
cohomology of ${\bb O} P^2$, i.e. that the second would probably be
a `composite' of the first. 

\vspace{3mm}
Imposing conventional Dirac quantization on the field ${\cal C}_7$
gives that these fields are classified topologically by a class 
$x \in H^8({\bb O} P^2 ;\Z)$, so that $x$ is represented in de Rham 
cohomology by $\frac{{\cal F}_8}{2\pi}$, 
$
x=\left[ \frac{{\cal F}_8}{2\pi} \right]$,
In analogy to the case in string theory \cite{Among} 
and M-theory \cite{Flux} \cite{Effective}, we consider the construction of the
partition function corresponding to  ${\cal C}_7$. This is done in terms of 
a theta function on $T=H^8\left( {\bb O} P^2 ; U(1) \right)$. However, since 
${\bb O} P^2$ has no torsion in cohomology, then $T$ will be the torus
\(
T= H^8({\bb O} P^2 ; \R)/ H^8({\bb O} P^2 ; \Z) \;.
\)
Furthermore, our construction requires introducing a function
\(
\Omega : H^8( {\bb O} P^2 ; \Z) \longrightarrow \Z_2\;,
\) 
which obeys the law
$
\Omega (x + y) = \Omega (x) \Omega (y) (-1)^{x \cdot y}$,
where $x \cdot y$ is the intersection pairing 
$\int_{{\bb O} P^2} x \cup y$ on ${\bb O} P^2$. The function $\Omega$
enters into the determination
of the line bundle ${\mathcal L}$ on $T$. The partition function
of the ${\cal C}_7$ field will then be a holomorphic section 
of ${\cal L}$.

\vspace{3mm}
The signature of ${\bb O} P^2$, which has dimension 16, is by definition the signature
of the quadratic form
$
H^8 ( {\bb O} P^2; \Q) \to \Q$, given by 
$
v \mapsto \langle v^2, [{\bb O} P^2] \rangle$,
and whose value is 1. 

\paragraph{The intersection form.}
For a manifold $M^{2n}$ of dimension $2n$, 
the universal coefficient theorem implies that 
\(
H_n(M^{2n}; \R) \cong H_n(M^{2n}) \otimes \R
\cong (H_n(M^{2n})/T_n) \otimes \R\;.
\)
Torsion elements do not affect the intersection number:
if $\alpha_n, \beta_n$ are torsion elements so that 
$r\alpha_n, s\beta_n \in H_n(M^{2n};\R)$, then
\(
\langle r\alpha_n, s\beta_n \rangle 
= rs \langle \alpha_n, \beta_n \rangle,
\)
so that the intersection forms over $\R$ and $\Z$ 
have the same matrix. Then $H_n (M^{2n};\R)$ has
a basis in which the intersection form has integer 
coefficients. Since the cup product is anti-commutative
then the intersection form is symmetric for even $n$ and
antisymmetric for odd $n$. The intersection form of
${\bb O} P^2$ is not even. This can be seen in two ways.
First that the signature of ${\bb O} P^2$, 
which is the signature 
of the intersection matrix of the middle cohomology of ${\bb O} P^2$,
is not zero. Second, the Steenrod operation 
$Sq^4k$ does not decompose in the similar way that
$Sq^{4k +2}$ does. In the latter case, the Adem relation
$Sq^{4k+2}=Sq^2 Sq^{4k} + Sq^1 Sq^{4k} Sq^1$ 
implies that $x_{4k+2}^2=Sq^{4k+2}x_{4k+2}=0$.

\vspace{3mm}
Now we look at mod 2 and integral bilinear forms. We have
\begin{prop}
\begin{enumerate}
\item The bilinear form
$
H^8({\bb O} P^2;\Z_2) \times H^8({\bb O} P^2;\Z_2) \rightarrow \Z_2$
over $\Z_2$
defined by 
$
(a_8 , a_8) \mapsto \int_{{\bb O} P^2} a_8 \cup a_8$
is given by $\int_{{\bb O} P^2} a_8 \cup w_8$.
\item The bilinear form over $\Z$, 
$H^8({\bb O} P^2;\Z) \times H^8({\bb O} P^2;\Z) \longrightarrow \Z$,
is an odd $\Z$-form. 
\end{enumerate}
\end{prop}
\proof
Consider the first part. Since $w_8^2=p_4$ mod 2 and
$w_{16}=3u^2=e$ mod 2, then the total Stiefel-Whitney class of ${\bb O} P^2$
is $w=1+u+u^2$, with coefficients of $u$ reduced mod 2 \cite{BH} (see 
equation (\ref{sw}).
The fact that the first seven Stiefel-Whitney classes of ${\bb O} P^2$ vanish
implies that
the Wu class $\nu ({\bb O} P^2)$ reduces
to the element $w_8({\bb O} P^2) \in H^8({\bb O} P^2; \Z_2)$ \cite{Hus}. Consequently, 
the Stiefel-Whitney class $w_8({\bb O} P^2)$ is characterized by the condition 
\cite{Fried}
\(
y_8 \cup y_8 = y_8 \cup w_8({\bb O} P^2)~ {\rm for~ any~} y_8 \in H^8({\bb O} P^2; \Z_2).
\)

\vspace{3mm}
Next consider the second part. In \cite{Fried} it 
was shown that, for a compact manifold $M^{16}$ admitting 
a ${\rm Spin}(9)$-structure, the quadratic form 
\(
H^8(M^{16};\Z)/{\rm Tor} \times H^8(M^{16};\Z)/{\rm Tor} 
\longrightarrow H^{16}(M^{16};\Z)
\)
is an even $\Z$-form if and only if $w_8(M^{16})=0$. Since
${\bb O} P^2$ has no torsion in cohomology, $H^{16}({\bb O} P^2;\Z)=\Z$,
and $w_8({\bb O} P^2)$ is nonzero, then the result follows immediately.
\endproof

In fact, we know that the value of the intersection form is given by the signature,
which is 1.

\subsection{Further terms and compatibility with other theories}
\label{sec:further}
\subsubsection{Kinetic terms}
\label{sec:kin}
We have not so far included any kinetic terms in the discussion. 
The main reason is that we do not know the nature of the resulting
`theory' and whether it will have such terms. If we take the proposal
in \cite{HS}, there are difficulties with the Einstein-Hilbert, i.e. the
gravitational kinetic, term because the obvious choice does
not give the correct term in bosonic string theory in twenty-six dimensions
upon dimensional reduction, but is off by a factor of $125/121$. This
is also linked with difficulties of finding coset symmetries 
\cite{Keur} \cite{LW}. Thus we exclude the gravitational terms from the 
discussion.
We go back to some of this in section \ref{compat}.
To some limited extent, we do consider the kinetic term for the M-theory
$C$-field provided this field lifts and provided that such a term does
in fact appear. 

\vspace{3mm}
Assuming a kinetic term for $G_4$, then the EOM would be
rationally 
\( 
d*_{27}G_4=\frac{1}{2} G_4 \wedge G_4 \wedge
Z_{16} + I_8 \wedge Z_{16},
\) 
where $*_{27}$ is the Hodge duality operator in 27 dimensions. 
The right hand side is a degree 24 differential form, whose class is
of the form
\(
\Theta_{24}^{\R}:=\left[ \frac{1}{2}G_4 \wedge G_4 + I_8\right] \wedge Z_{16}.
\label{form}
\)
As we have argued earlier, a term such as $Z_{16}$ can only
be a composite, i.e. a square of degree eight expressions, due
to the cohomology of ${\bb O} P^2$. 
We are interested in the integral lift of that degree 24 expression.
The term in brackets in (\ref{form}) has an integral lift given by
the class $\Theta_8$, defined in \cite{DFM},  as
$[\Theta_8(a)]_{\R}=\frac{1}{2}a_{\R}(a_{\R} - \lambda_{\R}) + 30 \widehat{A}_8$.
The integral lift of
$Z_{16}$ is just $u^2$ where $u$ is the generator of
$H^8({\bb O} P^2; \Z)$. Thus we have

\begin{prop}
The integral lift of $\Theta_{24}^{\R}$ is given by 
$
[\Theta_{24}]=[\Theta_8] \cup u^2
$.
\end{prop}
The study of this class, and further refinements
thereof, could be useful. 

\paragraph{Remark.}
Having $*_{27}G_4$ and $[\Theta_{24}]$ signals the appearance 
of 21-branes in the 27-dimensional theory. Requirement of 
decoupling of this brane from the membrane, so that a 
well-defined partition function can be constructed, gives that
the class $[\Theta_{24}]$ be trivial in cohomology, so that
the fields are cohomologically trivial on the brane. One obvious 
way to ensure this is to require triviality of $[\Theta_8]$. 
If we do not require this then we can find some other 
way to do this. We do not just set $u$ to zero. But we can
do something when reducing coefficients. 
Let $P_5^1$ be the Steenrod reduced power
operation 
$P_5^1: H^k({\bb O} P^2; \Z_5) \longrightarrow H^{k+8}({\bb O} P^2; \Z_5)$.
 Let
$\overline u$ be the generator $u$ with coefficients reduced mod 5.
In this case, for $k=8$, 
the action of $P_5^1$ is given by multiplication
with $5L_2$, where $L_2$ is the degree 8 term in the 
$L$-genus \cite{Inertia}. 
\( P_5^1{\overline u}=\frac{1}{9}(7p_2-p_1^2){\overline u}=
-2p_2{\overline u}=-2{\overline u}^2. 
\) 
This implies the following.

\noindent {\bf 1.} We can make $[\Theta_{24}]$ zero by imposing
the condition $P_5^1 \overline{u}=0$. This is analogous to the
mod 3 case in \cite{KSpin}.
 
\noindent {\bf 2.} For each
homeomorphism $\phi : {\bb O} P^2 \to {\bb O} P^2$, $\phi^* {\overline
u}={\overline u}$ \cite{BH}. Hence $\overline{u}$ is invariant
under continuous deformations of ${\bb O} P^2$.

\subsubsection{Compatibility with ten-dimensional superstring theories}
\label{sec:10}

We have looked at the proposed `theory' in twenty-seven 
dimensions in relation to 
M-theory in eleven dimensions. The question will now be whether
the structures we discussed are compatible with other known 
theories. Given that the 27-dimensional `theory' is proposed in 
such a way that it is by construction compatible with M-theory (as
we know it) then, since all five superstring theories in ten dimensions
are obtained from M-theory via dimensional reduction and/or dualities,
the 27-dimensional construction is compatible with these superstring
theories. We will actually reduce the $F_4-{\bb O} P^2$-bundle to ten dimensions
along the M-theory circle and check this explicitly.

\vspace{3mm}
We consider the ${\bb O} P^2$ bundle $M^{27}$ with structure group $F_4$. 
The transition functions on $Y^{11}$, with patches $U_i$ and $U_j$, 
will be
$
{g}_{ij} : {U}_i \cap {U}_i \longrightarrow {\rm Diff}({\bb O} P^2)$,
which are ${\rm Diff}({\bb O} P^2)$-valued (${\rm Diff}^+$ if orientation-preserving). 
If we take $Y^{11}$ to be the product
$X^{10} \times S^1$ and view the circle as the interval $[0, 1]$ with the
ends glued together then we can form the diagram
\(
\xymatrix{
{\bb O} P^2 
\ar[rr]^{=}
\ar[d]
&&
{\bb O} P^2
\ar[d]
\\
M^{27} 
\ar[rr]
\ar[d]
&&
\pi^* M^{27}
\ar[d]
\\
X^{10} \times S^1
&&
X^{10} \times [0, 1]
\ar[ll]^{\pi} \; .
}
\)
The bundle $\pi^* M^{27}$ is isomorphic to a bundle 
$\xi^{26} \times [0, 1]$ over $X^{10} \times [0,1]$. Gluing at
$[0,1]$ we get a map from $X^{10}$ to ${\rm Aut}(\xi^{26})$, the 
automorphism  group of the bundle $\xi$. Therefore,

\begin{prop}
From a bundle $M^{27}$ over $X^{10} \times S^1$ we get 

\noindent 1. a bundle $\xi^{26} \to X^{10}$ with fiber ${\bb O} P^2$ and structure
group $F_4$, and

\noindent 2. a gauge group element of $\xi^{26}$, i.e. a map $X^{10} \to {\rm Aut}(\xi^{26})$.
\end{prop}
If the bundle is trivial then the automorphims of $\xi^{26}$ will be the 
automorphisms of the fiber, i.e. $F_4$. A map from $X^{10}$ to $F_4$
might then be regarded as a classifying map for based loop bundles,
since $B \Omega F_4=F_4$. Thus, in this special case, we have 
an $F_4$ bundle and an $\Omega F_4$ bundle over $X^{10}$. This
is analogous to the case of $E_8$ \cite{Sloop}.

\vspace{3mm}
The diffeomorphism group above is very large and is not easy to 
work with. Instead we will invoke a condition that is familiar from Kaluza-Klein 
theory, namely to assume that the original bundle comes from a principal
$F_4$-bundle 
\(
\xymatrix{
F_4 
\ar[r]
&
P
\ar[r]
&
Y^{11},
}
\label{F4bun}
\)
so that we effectively consider the reduction of the structure group 
${\rm Diff}^+({\bb O} P^2)$ to the subgroup $F_4$, the isometry group of the
${\bb O} P^2$ fiber. This is analogous to the case when $Y^{11}$ itself is taken
as the total space of a circle bundle over $X^{10}$. A priori the structure group
is ${\rm Diff}^+(S^1)$, in which the transition functions are valued. Restricting
to $U(1) \subset {\rm Diff}^+(S^1)$, we get a principal circle bundle
$U(1)  
\to
Y^{11} \to X^{10}$. In fact, in this case, the reduction is always possible and 
no condition is required. Now we are presented with a situation which is 
analogous to having an $E_8$ bundle \cite{Flux} in eleven dimensions that
we are asking to reduce to ten dimensions. The result, analogously to 
the $E_8$ case \cite{AE} \cite{TDMW}, is 
\(
\begin{matrix}
F_4&\longrightarrow & P\cr
&&\dwn\cr
S^1 & \longrightarrow & Y^{11}\cr
&&\dwn\cr
& & X^{10}\cr
\end{matrix}
\hsp{.5}\Longrightarrow\hsp{.5}
\begin{matrix}
LF_4&\longrightarrow & Q\cr
&&\dwn\cr
& & X^{10}\;.\cr 
\end{matrix}
\)
The homotopy type of $F_4$ is identical to the homotopy type of $E_8$ 
in degrees less than eleven, and so rationally $F_4 \sim S^3$, 
$\Omega F_4 \sim S^4$, so that $LF_4 \sim S^3 \times S^4$. Thus, at the
rational level, we expect a degree three and a degree four class 
from the $LF_4$ bundle. At the integral level, since $F_4 \sim K(\Z, 3)$, 
then 
\(
LF_4 \sim K(\Z, 3) \times K(\Z, 4),~~~~~~{\rm deg} < 11. 
\)
This can be shown as follows.
We have $LF_4$ bundles which are classified by maps to $BLF_4$.
The sequence $\Omega X \to LX \to X$ for $X=BF_4$ gives
\(
F_4 \longrightarrow LBF_4 \longrightarrow BF_4\;.
\label{seq}
\)
Since $F_4$ is connected, then $LBF_4$ and $BLF_4$ are homotopy
equivalent.
We can then replace $LBF_4$ with $BLF_4$ in (\ref{se}).
Since $2$ and $3$ are the only torsion primes
for $F_4$, then for $p \geq 5$ the sequence 
\(
\xymatrix{
F_4 
\ar[rr]
&&
BLF_4 
\ar[rr]^{\rm ev}
&& 
BF_4 
\ar@/^1pc/[ll]^s
}
\label{se}
\)
splits on mod $p$ cohomology, so that
\(
H^*(BLF_4;\Z_p) \cong H^*(BF_4; \Z_p) \otimes
H^*(F_4 ;\Z_p),~~~~{p \geq 5}\; , 
\)
as algebras. At the torsion primes we use the Serre spectral sequence
 corresponding to the sequence (\ref{se}).
 From (\ref{bfz2}) we see for 
 $p=2$ that
in degrees $\leq 15$, 
\( 
H^*( BF_4;\Z_2)=\Z_2 \left[ x_4, Sq^2 x_4,
Sq^3 x_4\right] \; .
\label{bfz2 trun}
\) 
The differential $d$ acting on $x_4$ is zero because of the 
section $s$ in (\ref{se}).
From (\ref{bfz2 trun}), for $p=2$, and from (\ref{trun}), for $p=3$,
we see that all the generators are connected by cohomology 
operations, $Sq^i$ and $P^j$, respectively. Thus, since
$x_{i > 4}= \mathcal{O} x_4$, for some cohomology operation
$\mathcal{O}$, then all the differentials are zero. Thus the spectral
sequence collapses and the fibration is a product.  

\vspace{3mm}
The $LF_4$ bundle over $X^{10}$ is therefore a $K(\Z, 2) \times K(\Z, 3)$
bundle. The first factor, $K(\Z, 2)$ gives the NS field $H_3$ and
the second factor, $K(\Z, 3)$ gives the RR field ${\sf F}_4$ in ten dimensions.
Hence at the topological level, 
compatibility of $F_4$ with ten-dimensional type IIA is reduced to 
that of $E_8$, which follows from \cite{DMW} \cite{AE} \cite{TDMW}. 
The compatibility with type IIB, and hence with F-theory, also 
follows from T-duality as for the $E_8$ case \cite{E8T}. Therefore, 
we can give the following statement.

\begin{prop} Consider the ${\bb O} P^2$ bundle over $Y^{11}$ with structure 
group reduced to $F_4$ as above. Then

\noindent 1. The reduction of the $F_4$ bundle on the circle in $Y^{11}$ leads to 
an $LF_4$ bundle over $X^{10}$.

\noindent 2. At the topological level,
the ${\bb O} P^2$ bundle, with the above assumptions, is compatible 
with type II string theory. 
\label{thm:LF4}
\end{prop}

\subsubsection{Compatibility with the bosonic string}
\label{compat}

The question is whether the 27-dimensional structure is
compatible with the bosonic string theory in twenty-six dimensions, on 
$X^{26}$. We have addressed some aspects of this in section \ref{susy} 
in relation to fermions and supersymmetry, and so we consider 
other aspects in this section.
The form fields we have introduced, including $G_4$ from M-theory,
are all of dimensions that are multiples of 4. Since the bosonic string 
spectrum does not involve $G_4$ and the action does not obviously
get the topological terms that we introduced, then the relation between 
$M^{27}$ and $X^{26}$, if a dimensional reduction, could be 
a one-dimensional orbifold, 
\footnote{Alternatively, the relation between
the twenty-seven - and the twenty-six-dimensional theories could be 
more involved such as in the case of heterotic/type II duality.}
i.e. $S^1/\Z_2$, where we assume a $\Z_2$ 
parity on all form fields of degrees of the form $4k$ in such a way that
they disappear in the same way that $G_4$ gets killed in going from 
M-theory to the heterotic theory and also from the bosonic theory 
in 
\cite{HS} to twenty-six dimensions. Thus, the forms coming from the
${\bb O} P^2$ bundles can be made compatible with bosonic string theory. 

\vspace{3mm}
One difficulty with the proposal in \cite{HS} was raised in \cite{LW}, 
which is that the action does not support a coset symmetry that would
include the bosonic string theory. This was also observed in \cite{Keur}.
The question is whether our proposal can evade these objections.
In \cite{LW} the reduction was on tori, but ours is a coset space with
large and sparse homotopy cells. In \cite{Keur} the analysis was based on 
assumptions, such as Lorentz symmetry, that we do not know 
whether they hold for the higher-dimensional case, and the search was
made based on the classification of simple Lie algebras. It is possible 
that the higher structures will not be entirely described by such classical
notions (although of course we used some of these notions in our
own discussion). Furthermore, in both \cite{LW} and \cite{Keur} 
gravity was involved. The Einstein-Hilbert term in twenty-seven
dimensions does not give the correct term in twenty-six dimensions
\cite{HS}, and this is related to the lack of coset symmetry structure 
\cite{LW} mentioned above. We have not included the gravitational
kinetic terms in our discussion, mainly for this reason, but also 
because there is a possibility that the theory will not be of the 
the usual form. This was also raised in \cite{LW}. It is possible that
the theory will be nonlocal or topological. We cannot answer this 
in any definitive way here.

\vspace{3mm}
Thus, given the discussion about supersymmetry
at the end of section \ref{susy} and the above discussion, it would be 
desirable to find a compatibility diagram of the schematic form
\(
\xymatrix{
M^{27}
\ar[rr]^{?}
\ar[d]_{{\bb O}  P^2 {~\rm reduction}}
&&
X^{26}
\ar[d]^{\rm Lattice ~reduction}
\\
Y^{11}
\ar[rr]^{S^1~{\rm or~} S^1/\Z_2}_{\rm reduction}
&&
M^{10}\; .
}
\)
This requires further investigation but we have not immediately seen
an obstruction for this to hold.

\vspace{0.7cm}
{\bf \large Acknowledgements}

\vspace{2mm}
\noindent 
The author thanks the American Institute of Mathematics for hospitality and the 
`` Algebraic Topology and Physics" SQuaRE program participants 
for very useful discussions. 
The author would like to thank the Hausdorff Institute for Mathematics in Bonn
for hospitality and the organizers of the ``Geometry and Physics" Trimester Program
at HIM for the inspiring atmosphere during the writing of this paper.  Special 
thanks are due 
 Pierre Ramond for helpful remarks and encouragement and to Arthur Greenspoon for  
many useful editorial suggestions.

\section{Appendix: Some Properties of ${\bb O} P^2$}

In this appendix we summarize the topological 
properties of the Cayley plane ${\bb O} P^2$
which are useful for proving some of the results in the text.  

 \begin{enumerate}

\item {\bf Betti numbers:} The only nonzero Betti numbers are $b_0=1$,
$b_8=1$, $b_{16}=1$.

\item {\bf Integral cohomology:} The cohomology ring is 
$
H^*({\bb O} P^2; \Z)=\Z[u]/u^3$,
where $u \in H^8({\bb O} P^2; \Z)$ is the canonical 8-dimensional generator
coming from $S^8$. Thus $H^0\Z=H^8\Z=H^{16}\Z= \Z$ and $H^i=0$
otherwise. Note that there is no torsion in cohomology. Consider 
the last Hopf map $S^7 \longrightarrow S^{15} 
{\buildrel{f}\over \longrightarrow} S^8$. The spheres $S^7$ and 
$S^8$ are oriented, so that generators 
$a \in H^7(S^7;\Z)=\Z$ and $b \in H^8(S^8;\Z)$ 
can be specified. The mapping cone ${\cal C}(f)$ is ${\bb O} P^2$. The 
exactness of the cohomology long exact sequence corresponding to 
$f$ gives the isomorphisms 
\bea
\imath &:& H^{15}(S^{15};\Z) \buildrel \cong\over{\longrightarrow}
H^{16}({\bb O} P^2;\Z)
\nonumber\\
\jmath^* &:& H^{8}({\bb O} P^2;\Z) \buildrel \cong\over{\longrightarrow} 
H^8(S^8;\Z)\;.
\eea 
Let $a'=\imath (a) \in H^{16}({\bb O} P^2;\Z)$, and let $u \in H^8({\bb O} P^2;\Z)$
be the unique element such that $\jmath^*(u)=b$. Since $H^{16}({\bb O} P^2;\Z)=\Z$
then there exists a unique integer $H(f)$, the {\it Hopf invariant}, such that
$u \cup u = H(f) a'$. It is a classical result that this is 
equal to one. Therefore $a'=u^2$. This justifies the above claim about the
cohomology of ${\bb O} P^2$.

\item {\bf Euler class:} Let $u$ be a generator of $H^8({\bb O} P^2;\Z)$. The
Euler class of ${\bb O} P^2$ is $e=\pm 3 u^2$.

\item {\bf Pontrjagin classes}: The total tangential Pontrjagin class is given by
\cite{BH}
$
p(T {\bb O} P^2)= 1+ 6u +39u^2$,
so that the nonzero Pontrjagin classes are
$p_2=6u$, $p_4=39u^2$. Choosing that orientation which is defined by
$u^2$, the non-vanishing Pontrjagin numbers are $p_2^2[{\bb O} P^2]=36$,
$p_4[{\bb O} P^2]=39$.

\end{enumerate}


\end{document}